\documentclass[english]{article}
\usepackage{lmodern}
\usepackage[LGR,T1]{fontenc}
\usepackage[utf8]{inputenc}
\usepackage[a4paper]{geometry}
\usepackage{textcomp}
\usepackage{mathtools}
\usepackage{amsmath}
\usepackage{amssymb}
\usepackage{graphicx}
\usepackage{microtype}

\makeatletter

\ProvideTextCommand{\~}{LGR}[1]{\char126#1}

\usepackage[unicode=true, bookmarks=true,bookmarksnumbered=false,bookmarksopen=false, breaklinks=false,pdfborder={0 0 0},pdfborderstyle={},backref=false,colorlinks=false]{hyperref}
\hypersetup{pdftitle={Review and experimental verification of X-ray darkfield signal interpretations with respect to quantitative isotropic and anisotropic darkfield computed tomography}, pdfauthor={Jonas Graetz (Dittmann), Andreas Balles, Randolf Hanke, Simon Zabler}}
\newcommand{\appropto}{\mathrel{\vcenter{
  \offinterlineskip\halign{\hfil$##$\cr
    \propto\cr\noalign{\kern2pt}\sim\cr\noalign{\kern-2pt}}}}}
\usepackage{fancyhdr}
\makeatother

\usepackage{babel}
\begin{document}
\title{\date{}Review and experimental verification of X-ray darkfield signal
interpretations with respect to quantitative isotropic and anisotropic
darkfield computed tomography }
\author{J. Graetz,$^{1,2,}$\thanks{Corresponding author. \texttt{jonas.graetz@physik.uni-wuerzburg.de}}
~A. Balles,$^{1,2}$ ~R. Hanke,$^{1,2,3}$ ~S. Zabler$^{1,2}$\\
{\footnotesize{}}\\
{\footnotesize{}$^{1}$}\emph{\footnotesize{}Lehrstuhl für Röntgenmikroskopie,
Universität Würzburg, Josef-Martin-Weg 63, 97074 Würzburg, Germany}{\footnotesize{}}\\
{\footnotesize{}$^{2}$}\emph{\footnotesize{}Fraunhofer IIS, division
EZRT, Flugplatzstraße 75, 90768 Fürth / Josef-Martin-Weg 63, 97074
Würzburg}\\
{\footnotesize{}$^{3}$}\emph{\footnotesize{}Fraunhofer IZFP, Campus
E3 1, 66123 Saarbrücken, Germany}}
\maketitle
\begin{abstract}
Talbot(-Lau) interferometric X-ray darkfield imaging has, over the
past decade, gained substantial interest for its ability to provide
insights into a sample's microstructure below the imaging resolution
by means of ultra small angle scattering effects. Quantitative interpretations
of such images depend on models of the signal origination process
that relate the observable image contrast to underlying physical processes.
A review of such models is given here and their relation to the wave
optical derivations by Yashiro et al\@.\ and Lynch et al.\ as well
as to small angle X-ray scattering is discussed. Fresnel scaling is
introduced to explain the characteristic distance dependence observed
in cone beam geometries. Moreover, a model describing the anisotropic
signals of fibrous objects is derived.  The Yashiro-Lynch model is
experimentally verified both in radiographic and tomographic imaging
in a monochromatic synchrotron setting, considering both the effects
of material and positional dependence of the resulting darkfield contrast.
The effect of varying sample--detector distance on the darkfield
signal is shown to be non-negligible for tomographic imaging, yet
can be largely compensated for by symmetric acquisition trajectories.
The derived orientation dependence of the darkfield contrast of fibrous
materials both with respect to variations in autocorrelation width
and scattering cross section is experimentally validated using carbon
fiber reinforced rods.

\thispagestyle{fancy}
\end{abstract}

\section{Introduction}

A Talbot(-Lau) grating interferometer is a specific realization of
a shearing interferometer based on the Talbot-effect. It can be implemented
for the X-ray spectrum and, in addition to classic attenuation contrast,
further gives access to diffraction contrasts in the form of differential
phase shift information and ultra small angle X-ray scattering contrast
(cf.\ \cite{Cloetens1997,David2002,Momose2003,Weitkamp2005OptE,Pfeiffer2006}).
The latter is commonly referred to as ``darkfield contrast'' in
analogy to the respective scattering contrasts in other fields of
imaging and is of particular interest for its sensitivity to the unresolved
substructure of the sample. The unique advantages of the grating interferometer
with respect to other diffractive X-ray imaging techniques, such as
scanning small angle X-ray scattering (scanning SAXS), the crystal
analyzer based ``diffraction enhanced imaging'' or ``multiple image
radiography'' methods (cf.\ \cite{ZhongSayers2000,Olivo2002,Pagot2003,Wernick2003}),
and speckle-based imaging techniques (cf\@.\ the recent review by
Zdora \cite{Zdora2018}), are its ability to directly capture planar
images (in contrast to pixel or line scanning methods), its sensitivity
to micrometer scaled diffractive effects within centimeter sized fields
of view (in contrast to microscopy techniques) and finally its tolerance
to non-monochromatic X-rays and practical feasibility in laboratory
environments (as opposed to synchrotron facilities). The darkfield
contrast generated by Talbot interferometers (along with attenuation
and differential phase contrast) provides complementary information
both in non-destructive testing and life sciences due to its sensitivity
to sub-resolution structures such as micro cracks and porous or fibrous
matter. The initial interest on darkfield imaging arose from early
synchrotron experiments on mammography with monochromatic radiation
and crystal analyzers \cite{Johnston1996,Chapman1997}. With the introduction
of Talbot-Lau imaging to the laboratory by Pfeiffer et al.\ \cite{Pfeiffer2006,Pfeiffer2008},
many more examples have been shown. Applications of X-ray darkfield
imaging besides the characterization of micro-calcifications in mammography
\cite{Michel2013,Grandl2015} include imaging of lungs \cite{Yaroshenko2014,Velroyen2015,Gromann2017,Ludwig2019},
characterization of bone and dentin \cite{Potdevin2012,Momose2014,Jud2016,Jud2017,Horn2017}
as well as the analysis of general porous and fibrous materials or
microscopic defects in the field of non destructive testing \cite{Revol2011,Jerjen2013,Gresil2017}.
Other applications that have been shown include water transport in
cement \cite{Prade2016,Yang2018} or the monitoring of germinating
seeds \cite{Nielsen2017}. Due to its origin in the ultra small angle
scattering properties of a given sample, darkfield contrast also reflects
anisotropies in scattering and thereby allows to detect local orientations
within fibrous materials (cf.\ \cite{Jensen2010PMB,Jensen2010PRB,Potdevin2012,Bayer2012OE,Revol2013,Schaff2014,Hannesschlaeger2015}).

With respect to quantitative interpretations of the obtained images,
a solid understanding of the underlying contrast mechanisms is of
great importance, particularly in the light of long processing chains
beginning with phase stepping analysis (cf.\ \cite{Seifert2016,Kaeppler2017,DeMarco2018,Dittmann2018,Hashimoto2020}
and references therein) and ending in tensor valued volume reconstructions
of anisotropically scattering materials (cf.\ \cite{Bayer2014,Malecki2014,Vogel2015,Wieczorek2016,Dittmann2017,Graetz2019}).
The aim of the present article is to provide a unified view on the
numerous explanations on darkfield signal origination that have been
given in previous literature, and to provide experimental support
for the central results. The existing theories will further be extended
to cone beam geometries using the Fresnel scaling relation. Moreover,
a model for the description of generally oriented anisotropic scatterers
will be derived based on the concepts presented by Yashiro et al.\
\cite{Yashiro2010} and Lynch et al.\ \cite{Lynch2011}. Experimental
support will be given likewise.

\section{Talbot Interferometer}

\begin{figure}
\begin{centering}
\includegraphics[width=0.45\textwidth]{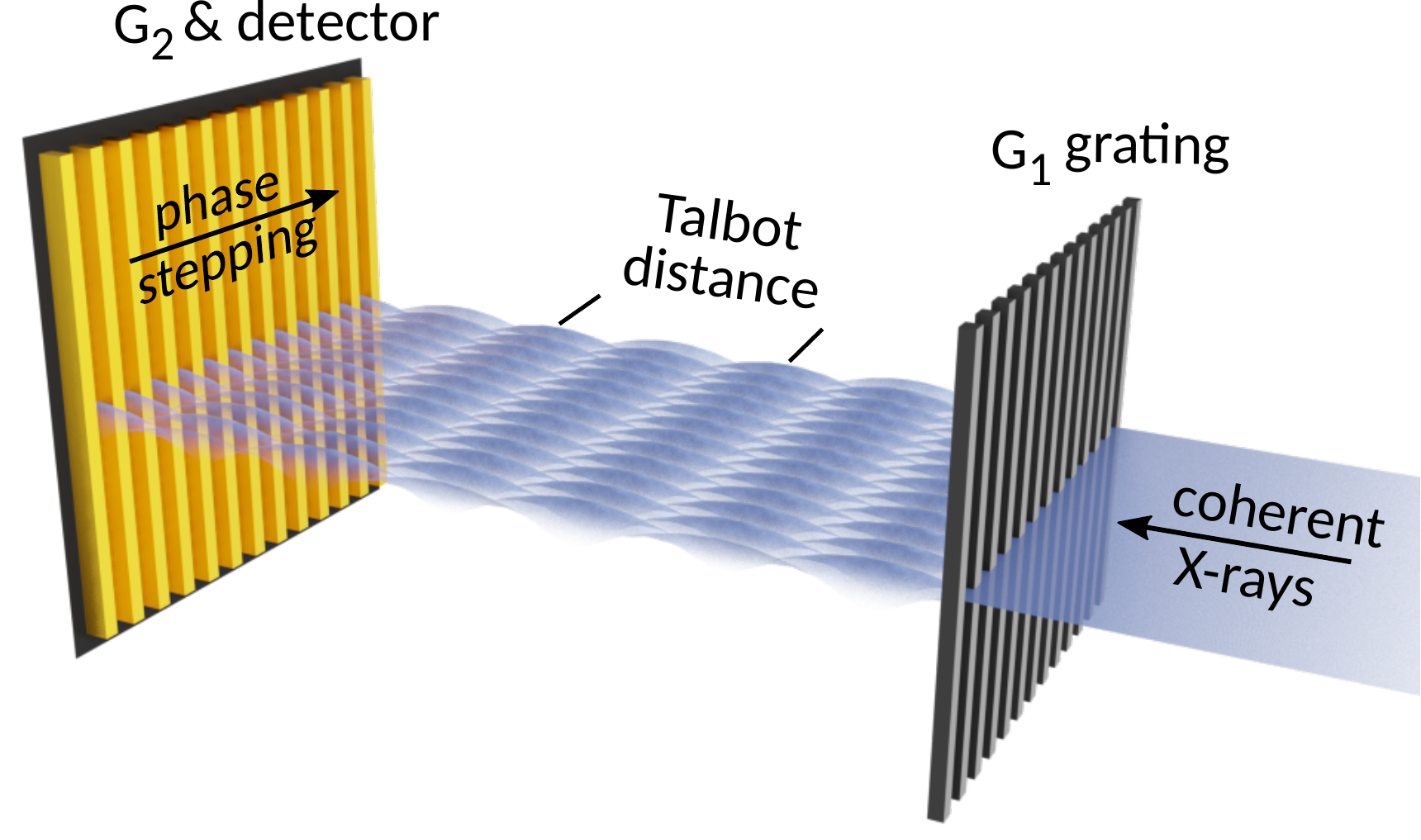}\hfill\includegraphics[width=0.535\textwidth]{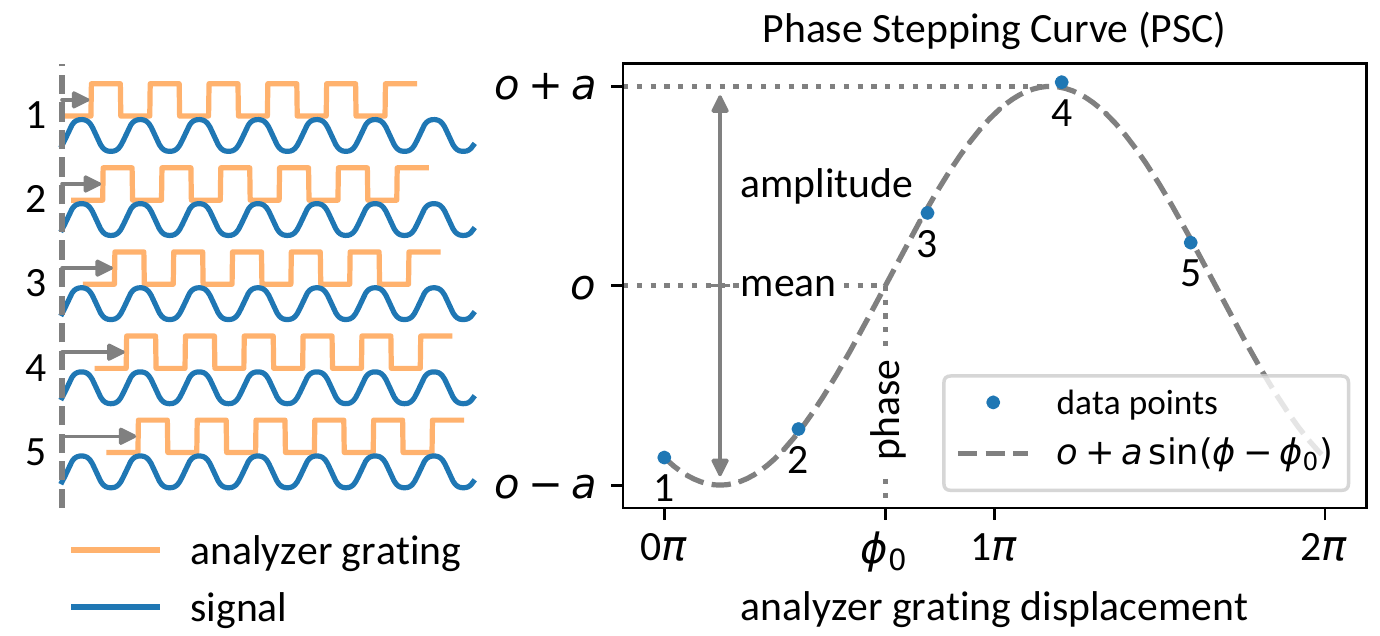}
\par\end{centering}
\centering{}\caption{\label{fig:Talbot-Interferometer}Sketch of a Talbot-interferometer
(left). Coherent X-rays are modulated by a periodic, phase-shifting
or absorbing grating G$_{\text{1}}$. Coherent diffraction within
the Fresnel-regime leads to a periodic reproduction (in intervals
of the Talbot distance) of the periodic amplitude modulation induced
by the G$_{\text{1}}$ grating (Talbot-effect). The patterned beam
can be analyzed by means of a period-matched absorption grating G$_{\text{2}}$
in front of an integrating detector. The detector pixel size is typically
at least one order of magnitude larger than the grating period. By
moving the analyzer grating G$_{\text{2}}$ in front of the detector
about one period in multiple steps, a phase stepping curve can be
acquired (cf.\ right panel. Stepping of G$_{\text{1}}$ will have
an equivalent effect). Wavefront changes due to a sample will result
in attenuation, deflection or blurring of the G$_{\text{1}}$-induced
beam modulation and results in corresponding changes of the phase
stepping curve. Samples may be placed between G$_{\text{1}}$ and
G$_{\text{2}}$ or in front of G$_{\text{1}}$. Experimental realizations
often use an additional absorption grating G$_{\text{0}}$ before
G$_{\text{1}}$, structuring common laboratory X-ray sources into
multiple narrow slit sources of sufficient coherence (Talbot-Lau interferometer).}
\end{figure}
Generally, an X-ray Talbot-Lau interferometer consists of three micrometer-pitched
gratings, of which the first is placed close to the X-ray source,
shaping it into multiple small slit sources in order to increase coherence
of the emitted radiation. It can be omitted for sufficiently coherent
sources such as synchrotron radiation or microfocus X-ray tubes (Talbot
interferometer, cf.\ Fig.~\ref{fig:Talbot-Interferometer}). The
second grating imposes a periodic phase or amplitude modulation. Although
the coherent wavefront is subject to interference while propagating,
its periodic intensity modulation is restored at specific distances
characteristic to the grating pitch and X-ray energy (Talbot effect).
At these distances, it can be analyzed by means of a third grating
with matching periodicity in combination with an X-ray detector placed
behind it (having a pixel size considerably larger than the grating
period). When the structured beam is perturbed by a sample, three
effects can be observed: As for classic X-ray imaging, the intensity
may be diminished due to absorption. Moreover, the periodic pattern
may be reduced in contrast (visibility) due to scattering, and shifted
in phase due to refraction. The effects of a sample can be analyzed
by comparison of the phase stepping curves for the perturbed (by the
sample) and unperturbed beam.

\section{Isotropic Darkfield Contrast\label{subsec:DF-Interpretation}}

Analog to the characterization of the modulation transfer properties
of an optical system, visibility or contrast reduction of the structured
beam (i.e., the reduction of the first harmonic Fourier component
normalized to the zeroth as e.g.\ defined by \cite{Pfeiffer2008})
can generally be modeled as Gaussian blurring of the reference sinusoid
profile. The underlying conception is a convolution of the incident
intensity distribution with an effective point spread function or
scattering profile of a sample. Normalizing the Gaussian to the ratio
of transmitted intensity $t\in[0,1]$ (i.e., also accounting for attenuation
or general reduction of intensity), the following convolution kernel
can be stated:
\begin{equation}
\frac{t}{\sqrt{2\pi}\sigma_{\phi}}e^{-\frac{1}{2}\frac{\Delta\phi^{2}}{\sigma_{\phi}^{2}}}\:,
\end{equation}
with $\sigma_{\phi}$ being its standard deviation in units of radians
and $\Delta\phi$ the phase with respect to the periodic irradiation
profile (not to be confused with a scattering angle). The width $\sigma_{\phi}$
will later be expressed in terms of geometric parameters based on
absolute scales of the instrument. Yet to begin with, the present
formulation will be convenient. 

By explicitly solving the convolution integral over a generic sinusoid
reference profile $o_{\mathrm{ref}}+a_{\mathrm{ref}}\cos(\phi)$:
\begin{equation}
\frac{t}{\sqrt{2\pi}\sigma_{\phi}}\int_{-\infty}^{+\infty}\text{d}\Delta\phi\,(o_{\mathrm{ref}}+a_{\mathrm{ref}}\cos(\Delta\phi-\phi))\,e^{-\frac{1}{2}\frac{\Delta\phi^{2}}{\sigma_{\phi}^{2}}}=\underbrace{t\,o_{\mathrm{ref}}}_{=o_{\mathrm{smp}}}+\,\underbrace{t\,a_{\mathrm{ref}}\,e^{-\frac{\sigma_{\phi}^{2}}{2}}}_{=a_{\mathrm{smp}}}\cos(\phi)
\end{equation}
and identifying the changed mean intensity and amplitude parameters
$o_{\mathrm{smp}}$ and $a_{\mathrm{smp}}$ on the right hand side,
the sample-induced visibility reduction $v,$ defined as $v=v_{\mathrm{ref}}/v_{\mathrm{smp}}=\frac{a_{\mathrm{smp}}/o_{\mathrm{smp}}}{a_{\mathrm{ref}}/o_{\mathrm{ref}}}$,
can be directly identified: 
\begin{equation}
v=e^{-\frac{\sigma_{\phi}^{2}}{2}}\:.\label{eq:DF-v-exp-sigma2}
\end{equation}
Due to the orthogonality of the Fourier basis, this result also applies
to non-sinusoidal phase stepping curves and generally describes the
contrast visibility of a particular harmonic of the periodic pattern
implicitly selected by the respective definition of $\phi$ with respect
to spatial dimensions, which will be concretized in the following.

\subsection{Linear diffusion interpretation \label{subsec:DF-diffusion-model}}

Given the period $T_{\mathrm{G2}}$ of the Talbot interferometer's
analyzer grating, the sample induced blurring width $\sigma_{\phi}$
of the first harmonic can be expressed in spatial units (indicated
by the subscript $x$):
\begin{equation}
\sigma_{x}=\frac{T_{\mathrm{G2}}}{2\pi}\sigma_{\phi}\:.
\end{equation}
When further arguing that the specific width $\sigma_{x}$ at the
location of the analyzer grating arises due to a beam divergence caused
by the sample at distance $d$, the respective divergence may as well
be characterized by an angular standard deviation $\sigma_{\theta}$.
Using the small-angle approximation $\Delta\theta\approx\tan\Delta\theta=\frac{\Delta x}{d}$
for $d\gg\Delta x$, we find:
\begin{align}
\sigma_{\theta} & =\frac{\sigma_{x}}{d}=\frac{T_{\mathrm{G2}}}{d}\frac{\sigma_{\phi}}{2\pi}\nonumber \\
 & =\frac{T_{\mathrm{G2}}}{2\pi\,d}\sqrt{-2\ln(v)}\:.\label{eq:DF-diffusion-sigma-theta}
\end{align}
Within the model of angular beam diffusion and geometric propagation
onto the analyzer grating, $\sigma_{\theta}$ represents an invariant
property of the sample (at a given X-ray energy), while $v$, $\sigma_{\phi}$
and $\sigma_{x}$ include properties of the instrument.

This geometric interpretation has e.g.\ been given by Wang et al.\
(2009 \cite{Wang2009}) to describe the origination of darkfield contrast,
and Bech and Grünzweig et al.\ (2010/2013 \cite{Bech2010,Gruenzweig2013})
introduced this concept as ``linear diffusion coefficient''
\begin{equation}
\epsilon=\frac{\sigma_{\theta}^{2}}{\Delta z}=-\frac{1}{\Delta z}\frac{T_{\mathrm{G2}}^{2}}{2\pi^{2}d^{2}}\ln(v)\:,\label{eq:DF-linear-diffusion-coeff}
\end{equation}
further normalizing $\sigma_{\theta}^{2}$ to the sample thickness
$\Delta z$ in order to obtain a thickness independent material constant
analog to the classic ``linear absorption coefficient'' $\mu$,
such that:
\begin{equation}
-\ln\left(v(x,y)\right)\propto\int\!\mathrm{d}z\,\epsilon(x,y,z)\:,
\end{equation}
with $z$ denoting the optical axis and $x,y$ being the planar projection
image coordinates.

Anticipating the following Sections, it shall be noted here that the
angular diffusion interpretation as presented here only holds in a
parallel beam geometry and in the limit of sufficiently large and
smooth scattering structures. The critical length scale depends on
the interferometer parameters, as will be explained in the following
(cf.\ Sections \ref{subsec:DF-abinitio} and \ref{subsec:DF-diffusion-approx}).
Additional geometric scaling effects beyond the trigonometric relation
$\sigma_{x}\approx d\sigma_{\theta}$ need to be explicitly accounted
for in the case of a Talbot interferometer in cone beam geometry,
i.e., in the typical laboratory use case (cf.\ Secion~\ref{subsec:DF-Fresnel-scaling}).

\subsection{Ab initio derivation of visibility reduction\label{subsec:DF-abinitio}}

Yashiro et al.\ and Lynch et al.\ (2010--2011 \cite{Yashiro2010,Lynch2011})
derive the origination of darkfield contrast (or visibility reduction)
starting from a first principles approach, explicitly modeling spatially
varying refractive indices of grating and sample. By coherent propagation
of the complex wavefront to the location of the analyzer grating within
the optical Fresnel regime and subsequent Fourier analysis (analog
to the effect of the analyzer grating), a complete model of the Talbot
imaging process in a parallel beam geometry is obtained. While Yashiro
et al.\ considered the sample to be placed in front of the Talbot
interferometer, Lynch et al.\ considered the case of the sample placed
between the modulator and analyzer grating. Both arrive at the same
conclusions employing a statistic model of the sample's refractive
properties on the scale below the system's spatial resolution. More
specifically, the sample is characterized by its total phase shift
$\Phi$ along the optical axis (its absorbing properties are treated
separately), which varies in the perpendicular $x,y$ plane (the imaging
plane), i.e., $\Phi=\Phi(x,y)$. Decomposing $\Phi(x,y)$ into low-
and high-frequency (smooth and fine) components $\Phi_{\mathrm{s}}(x,y)$
and $\Phi_{\mathrm{f}}(x,y)$, the darkfield contrast can be attributed
to the high frequency component $\Phi_{\mathrm{f}}$. High frequency
variations in absorption are neglected. The lower cutoff frequency
of $\Phi_{\mathrm{f}}$ is determined by the spatial resolution of
the imaging system, i.e., by its effective point spread width.

The final result given by both Yashiro and Lynch (although using different
notations) relates the visibility $v$ to the autocorrelation of the
high frequency part $\Phi_{\mathrm{f}}$ of the sample's phase shifting
properties $\Phi$:
\begin{equation}
v=e^{-\sigma_{\Phi_{\mathrm{f}}}^{2}(1-\gamma(\xi))},\label{eq:yashiro-visibility}
\end{equation}
with $\gamma(\xi)$ being the normalized autocorrelation function
(i.e., $\gamma(0)=1$), $\sigma_{\Phi_{\mathrm{f}}}^{2}$ the variance
of $\Phi_{\mathrm{f}}$ and $\xi$ the correlation distance. $\gamma(\xi)$
and $\sigma_{\Phi_{\mathrm{f}}}^{2}$ are to be understood as respective
mean properties of $\Phi_{\mathrm{f}}$ on the spatial scale of the
imaging system's point spread width at a given location $(x,y)$.
The explicit $x,y$ dependence has been omitted here for improved
readability. The one dimensional autocorrelation parametrized by $\xi$
is performed along the interferometer's sensitivity direction, i.e.,
perpendicular to the interferometer grating bars. This for most setups
is the horizontal or $x$ direction.

The correlation distance $\xi$ is determined by the X-ray wavelength
$\lambda$, the analyzer grating period $T_{\mathrm{G2}}$ and the
distance $d$ between sample and analyzer grating \cite{Yashiro2010,Lynch2011}:
\begin{equation}
\xi=\frac{\lambda}{T_{\mathrm{G2}}}d\,.\label{eq:correlation-length}
\end{equation}
The correlation distance can thus be easily tuned by varying $d$,
i.e., by moving the sample closer to or further from the analyzer
grating. The system constants $\lambda$ and $T_{\mathrm{G2}}$ (and
the maximum possible distance $d$) determine the accessible order
of magnitude of $\xi$, which typically ranges at the micrometer level
for most X-ray Talbot interferometers.

Although Yashiro and Lynch did not explicitly consider the case of
$d$ being larger than the G$_{\text{1}}$--G$_{\text{2}}$ distance
(i.e., the sample was assumed to be either directly in front of G$_{\text{1}}$,
or between G$_{\text{1}}$ and G$_{\text{2}}$), the employed Fresnel
propagator formalism (cf.\ \cite{Paganin}) doesn't give reason to
expect this situation to be fundamentally different.

The distinction between effective cross section $\sigma_{\Phi_{\mathrm{f}}}^{2}$
and autocorrelation $\gamma(\xi)$ can also be interpreted as a representation
of orthogonal sample properties. While $\gamma(\xi)$ is characteristic
to the structure perpendicular to the optical axis, $\sigma_{\Phi_{\mathrm{f}}}^{2}$
characterizes the sample along the optical axis. In particular, $\sigma_{\Phi_{\mathrm{f}}}^{2}$
will scale with sample thickness, while $\gamma(\xi)$ depends on
the sample's characteristic structure.

\subsection{Relation to the linear diffusion interpretation\label{subsec:DF-diffusion-approx}}

The geometric beam diffusion interpretation given by Wang, Bech and
Grünzweig et al.\ (2009--2013 \cite{Wang2009,Bech2010,Gruenzweig2013})
can be found as a special case of the more general result $v=\exp(-\sigma_{\Phi_{\mathrm{f}}}^{2}(1-\gamma(\xi))$
given by Yashiro and Lynch when considering the limit of smooth convex
(spheroid) scattering structures of diameter $D\gg\xi$. The autocorrelation
function $\gamma$ can then be approximated by $\exp(-\frac{1}{2}\frac{\xi^{2}}{(D/3)^{2}})$
(cf.\ \cite{Prade2015} and references therein):
\begin{align}
\gamma_{D}(\xi) & \approx e^{-\frac{1}{2}\frac{\xi^{2}}{(D/3)^{2}}}\label{eq:gaussian-sphere-corr-approx}\\
\text{(parabolic approx.)} & \approx1-\frac{1}{2}\frac{\xi^{2}}{(D/3)^{2}}\quad\text{for }D\gg\xi\:,\nonumber 
\end{align}
such that
\begin{align}
v & \approx\exp\left(-\frac{1}{2}\frac{\sigma_{\Phi_{\mathrm{f}}}^{2}\xi^{2}}{(D/3)^{2}}\right)=\exp\biggl(-\frac{1}{2}\underbrace{\left(\frac{\sigma_{\Phi_{\mathrm{f}}}}{D/3}\right)^{2}\left(\frac{\lambda}{T_{\mathrm{G2}}}\right)^{2}d^{2}}_{=\sigma_{\phi}^{2}}\biggr)\:.\label{eq:yashiro-sigma-approx}
\end{align}
In this limit, the linear distance dependence of the blurring width
$\sigma_{\phi}$ on the sample distance $d$ as assumed in the angular
diffusion interpretation is reproduced. By substitution of Eq.~\ref{eq:yashiro-sigma-approx}
into Eq.~\ref{eq:DF-linear-diffusion-coeff}, the following identity
results:
\begin{equation}
\epsilon\approx\frac{1}{\Delta z}\left(\frac{\lambda}{2\pi}\frac{\sigma_{\Phi_{\mathrm{f}}}}{D/3}\right)^{2}
\end{equation}
The linear diffusion model becomes inaccurate as soon as $\gamma$
deviates from a parabolic approximation, which either is the case
for too large values of $\xi/D$ (i.e., for too small diameters $D$),
or when $\gamma$ actually is not parabolic even for small $\xi/D$.
The former case applies for too small scattering particles, while
the latter case applies for non-smooth shaped structures. 

While the geometric beam diffusion interpretation was originally proposed
independent of the particular imaging geometry (parallel beam versus
cone beam), it's relation to the more rigorous Yashiro-Lynch theory
suggests that additional geometric scaling effects as predicted by
coherent optics need to be taken into account when considering laboratory
instruments operating in cone beam geometry.

\subsection{Magnification and Fresnel scaling in cone beam geometry\label{subsec:DF-Fresnel-scaling}}

The derivations by Yashiro and Lynch, as well as the numeric simulations
by Malecki et al.\ \cite{Malecki2012} confirming their results,
were explicitly performed in a planar illumination context, which
in experimental terms is commonly given only at synchrotron facilities.
Laboratory setups operate at moderate distances from an X-ray point
source as compared to the extent of the illuminated field of view
and therefore exhibit non-negligible changes in geometric magnification
for different positions along the optical axis. However, by means
of the Fresnel scaling theorem, all results obtained for plane wave
illumination within the Fresnel approximation can be directly transferred
to a cone beam or point source illumination scenario by geometric
scaling of all dimensions, i.e.\ \cite{Paganin}:
\begin{align}
I^{(\mathrm{SOD})}(x,y,d) & =M^{-2}I^{(\infty)}(\frac{x}{M},\frac{y}{M},\frac{d}{M})\\
M & =\frac{\mathrm{SDD}}{\mathrm{SOD}}=\frac{\mathrm{SDD}}{\mathrm{SDD}-d}\:,\label{eq:fresnel-scaling-M}
\end{align}
with $M$ being the geometric magnification factor defined by the
ratio of source--detector distance $\mathrm{SDD}$ (assumed to be
equivalent to the source--G$_{\text{2}}$ distance) and source--object
distance $\mathrm{SOD}$. The optical axis is oriented along the $z$-axis,
$x$ and $y$ are the orthogonal image plane coordinates. $I^{(\mathrm{SOD})}$
is the intensity distribution at the detector (placed directly behind
the analyzer grating G$_{\text{2}}$ at distance $d$ from the sample)
in the case of a source placed at distance $\mathrm{SOD}$. $I^{(\infty)}$
is the intensity pattern for the case of an infinitely distant source
(i.e., the case of plane wave illumination). The Fresnel scaling theorem
essentially states that classic geometric magnification based on the
intercept theorem applies also to coherent wave propagation within
the Fresnel regime, affecting both the plane perpendicular to the
optical axis as well as propagation distances along the optical axis.

While this scaling relation is taken into account in the design of
laboratory type Talbot-Lau interferometers by scaling the position
and period of the modulating G$_{\text{1}}$ grating appropriately
with respect to the analyzer grating G$_{\text{2}}$ \cite{Weitkamp2005OptE,Weitkamp2006,Donath2009},
it has, to the author's knowledge, not yet been regarded in models
of the distance dependence of the visibility or darkfield contrast.

When consequently applying the above scaling relation to the correlation
distance $\xi$ (cf.\ Eq.~\ref{eq:correlation-length}), the effective
correlation distance in terms of the actual sample dimensions becomes,
for a cone beam (CB) geometry, 
\begin{equation}
\xi_{\mathrm{CB}}=\xi/M=\underbrace{\frac{\lambda}{T_{\mathrm{G2}}}d}_{\xi}\,(\underbrace{1-\frac{d}{\mathrm{SDD}}}_{=1/M(d)})\:.\label{eq:fresnel-scaled-xi}
\end{equation}
This result can be either interpreted as a consequence of scaled sampling
from the sample's phase $\Phi(x,y)$ according to the Fresnel scaling
theorem, or, more intuitively, as the consequence of the geometric
magnification of the sample at the location of the analyzer grating,
which serves as the reference scale for all other (downscaled) planes
along the optical axis (cf.\ the definition of $M$, Eq.~\ref{eq:fresnel-scaling-M}).

This also has implications on the angular beam diffusion model (cf.\
Sections \ref{subsec:DF-diffusion-model} and \ref{subsec:DF-diffusion-approx})
for the origination of darkfield contrast. By its conception of diffused
incident beams, geometric scaling was assumed to arise, independent
of the actual system geometry, only due to the trigonometric relation
between diffusion angle $\sigma_{\theta}$ and sample--detector distance
$d$. When considering the beam diffusion model as an approximation
of the Yashiro-Lynch theory in the case of large sample structures
in relation to $\xi$, as was discussed in Section~\ref{subsec:DF-diffusion-approx},
the following modified relation is found though:
\begin{equation}
v\approx\exp\left(-\frac{1}{2}\frac{\sigma_{\Phi_{\mathrm{f}}}^{2}\xi_{\mathrm{CB}}^{2}}{(D/3)^{2}}\right)=\exp\biggl(-\frac{1}{2}\underbrace{\left(\frac{\sigma_{\Phi_{\mathrm{f}}}}{D/3}\right)^{2}\left(\frac{\lambda}{T_{\mathrm{G2}}}\right)^{2}\left(\frac{d}{M(d)}\right)^{2}}_{\smash[b]{=\sigma_{\phi}^{2}}}\biggr)\:,
\end{equation}
i.e., $\sigma_{\theta}\propto\frac{d}{M(d)}$ rather than $\propto d$.
Consequences of the specific form of $\xi_{\mathrm{CB}}$ are its
symmetry about $d=\mathrm{SDD}/2$ and its maximum 
\begin{equation}
\xi_{\mathrm{CB,\max}}=\frac{\lambda}{T_{\mathrm{G2}}}\frac{\mathrm{SDD}}{4}
\end{equation}
at $d=\mathrm{SDD}/2$.

\begin{figure}
\centering{}\includegraphics[width=1\textwidth]{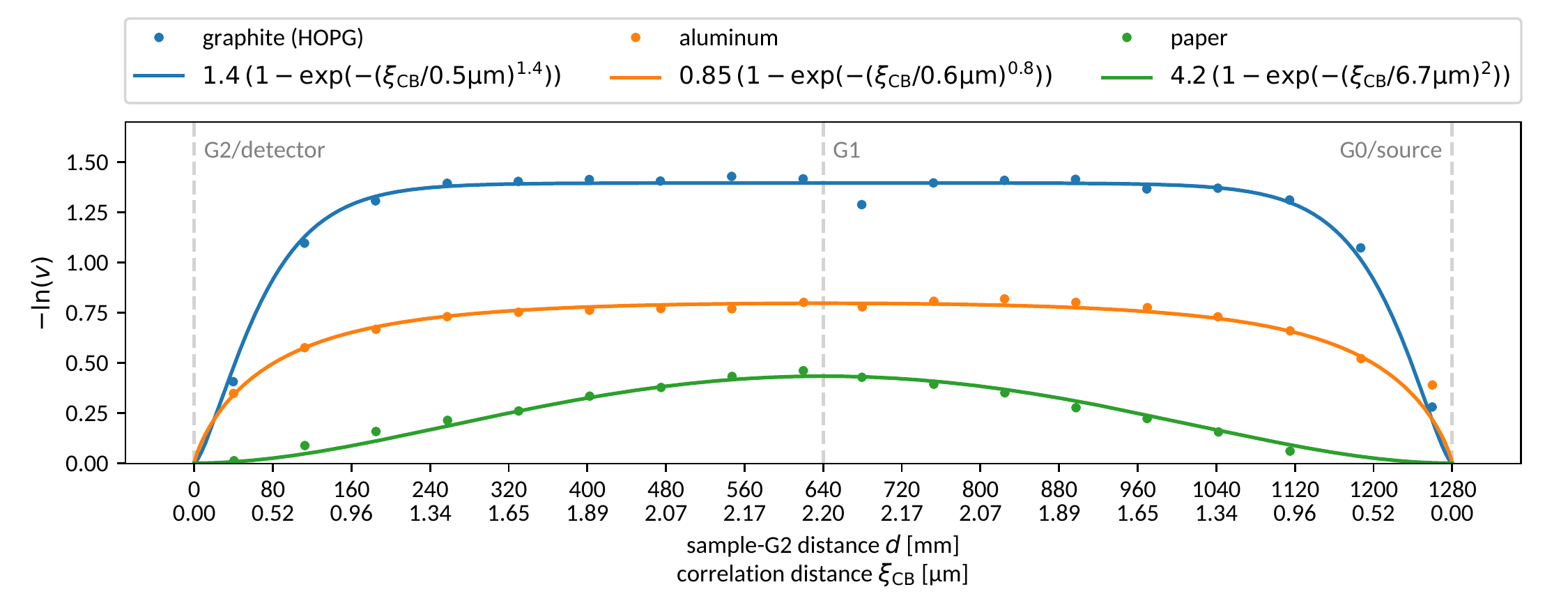}\caption{\label{fig:fresnel-scaling}Comparison of experimental data acquired
in cone beam geometry by M.\ Chabior \cite{Chabior2011} with the
Yashiro-Lynch model under consideration of Fresnel scaling (cf.\
Eq.~\ref{eq:fresnel-scaled-xi}) and a generic exponential autocorrelation
model (cf.\ Eq.~\ref{eq:exp-autocorr-model}).}
\end{figure}
Experimental evidence can be found e.g.\ in the Thesis of M.\ Chabior
\cite{Chabior2011}, who indeed finds, for a paper sample in a cone
beam setup, a deviation from the quadratic distance dependence of
$\ln(v)$ as expected by the linear diffusion model (cf.\ Section~\ref{subsec:DF-diffusion-model}).
The severe deviations found in the same work for aluminum and graphite
samples can on the other hand be explained by an additional violation
of the implicit $D\ll\xi$ assumption (cf.\ Section~\ref{subsec:DF-diffusion-approx}).
Eq.~\ref{eq:fresnel-scaled-xi} further reproduces the symmetry in
the distance dependence about the midpoint between source and G$_{\text{2}}$
that was experimentally shown e.g.\ by Chabior \cite{Chabior2011}
and Prade et al.\ \cite{Prade2015}. This symmetry has previously
been associated with a geometric argument given by Donath et al\@.\
\cite{Donath2009}, which was also adopted by Strobl \cite{Strobl2014}.
Figure~\ref{fig:fresnel-scaling} reproduces the data published by
Chabior \cite{Chabior2011} (using an interferometer designed to 45keV
with an analyzer grating period $T_{\mathrm{G2}}$ of 4\textmu m,
i.e., $\lambda/T_{\mathrm{G2}}\approx6.89\times10^{-6}$) and compares
it to the Yashiro-Lynch model under consideration of Fresnel scaling
(cf.\ Eq.~\ref{eq:fresnel-scaled-xi}) and the generic autocorrelation
model
\begin{equation}
\gamma(\xi_{\mathrm{CB}})=e^{-(\xi_{\mathrm{CB}}/L)^{2H}}\label{eq:exp-autocorr-model}
\end{equation}
also used in \cite{Yashiro2010}, with $L$ being a characteristic
correlation length and the Hurst exponent $H$ characterizing the
shape of the autocorrelation function and in particular allowing for
both convex (sharp) and concave (smooth) autocorrelation profiles.

Besides the correlation distance $\xi$, also the effective integration
area of a detector pixel with respect to the sample dimensions is
subject to geometric scaling. As the autocorrelation function $\gamma(\xi_{\mathrm{CB}})$
represents the mean statistical properties of the sample's phase shifting
properties below the imaging resolution (cf.\ Section~\ref{subsec:DF-abinitio}),
$\gamma$ itself will be subject to change whenever the the imaging
resolution (i.e., the pixel integration area) crosses characteristic
lengths scales of the sample's structure. This effect was experimentally
observed by Koenig et al.\ \cite{Koenig2016}, who further consistently
find a dependence also on the X-ray focal spot size, which likewise
affects the imaging resolution.

\subsection{Relation to Small Angle Scattering\label{subsec:DF-SAXS}}

Small angle scattering (SAS) techniques, or more specifically, small
angle X-ray scattering (SAXS), sense a sample's differential scattering
cross section (its scattering angle distribution) usually by means
of illuminating it with a collimated X-ray pencil beam. A planar detector
array placed behind the sample at a sufficient distance captures the
scattered radiation. By application of fundamental principles from
optics and Fourier analysis, the observable two-dimensional diffraction
pattern is commonly shown to correspond to the Fourier transform of
the auto correlation function of the sample's scattering length density
(cf.\ e.g.\ the textbook by da Sivia \cite{Sivia}). As the latter
is, on a larger scale, related to the sample's mass density, the observed
pattern can be, more qualitatively, understood as the Fourier transform
of the autocorrelation function of a sample's micro- or mesoscopic
structure. The resolution and range of sampled correlation distances
depends on the sensed angular range, with smaller scattering angles
corresponding to larger correlation distances, i.e., to larger structure
scales. 

While SAXS commonly addresses length scales in the $10^{1}$ to $10^{2}$
nanometer range, the signal captured by X-ray Talbot interferometers
corresponds to ultra small angle X-ray scattering (USAXS) typically
related to lengths scales on the micrometer scale (cf.\ Eqs.\ \ref{eq:yashiro-visibility}
and \ref{eq:correlation-length}). An important practical difference
between SAXS and USAXS with respect to data analysis and interpretation
emerges from the relation to the incident radiation: while the larger
scattering angles in SAXS generate signals well separated from the
incident radiation, USAXS analyses require explicit consideration
of contributions of the original beam.

Following the convolution concept of darkfield contrast origination,
Modregger et al.\ infer the effective point spread function (PSF)
of a sample by means of actual deconvolution of the phase stepping
curves (PSC) acquired with and without sample \cite{Modregger2012PRL}.
The result is directly interpreted, in analogy to classic SAXS, as
differential scattering cross section and Fourier transform of the
sample's autocorrelation function \cite{Modregger2014PRL,Modregger2017PRL}.

Another analogy between small angle scattering theory and darkfield
imaging was drawn by Strobl \cite{Strobl2014} and Prade et al.\
\cite{Prade2015}, who model an incoherent superposition of scattered
and unscattered fractions of the incident radiation in order to arrive
at an expression equivalent to the wave optical results by Yashiro
and Lynch (cf.\ Eq.~\ref{eq:yashiro-visibility}) in parallel beam
geometry. An experimental comparison of the latter model (and consequently
also of the Yashiro-Lynch model) to SAXS is given by Gkoumas et al.\
\cite{Gkoumas2016}, who investigate contributions of the structure
factor to the autocorrelation function -- as expected in classic
scattering experiments -- for dense sphere suspensions in a parallel
beam synchrotron setting.

The PSF deconvolution approach is, in terms of Fourier analysis, equivalent
to the evaluation of higher order Fourier components of the PSCs,
of which commonly only the zeroth (mean attenuation) and the first
(mean attenuation and visibility reduction) are evaluated. I.e., when
directly interpreting the point spread function (or convolution kernel)
as a SAXS pattern, the respective Fourier coefficients $q_{m}$ of
the phase stepping curves with and without sample are expected to
be related by an autocorrelation function $\gamma'(m\xi)$:
\begin{equation}
|q_{m}^{(\mathrm{PSC_{sample}})}|\propto\gamma'(m\xi)\,|q_{m}^{(\mathrm{PSC_{reference}})}|
\end{equation}
with $m$ enumerating the harmonics. The prime explicitly distinguishes
it from the autocorrelation function $\gamma$ as defined by Yashiro
and Lynch (and Strobl and Prade), although $\gamma$ and $\gamma'$
are expected to be closely related. When directly comparing $\gamma'$
with the Yashiro-Lynch theory for a thin sample and $m=1$, we find
(assuming $1\gg\sigma_{\Phi_{\mathrm{f}}}^{2}(1-\gamma(m\xi))\ge0$):
\begin{align*}
\gamma'(\xi) & \:\hat{=}\:e^{-\sigma_{\Phi_{\mathrm{f}}}^{2}(1-\gamma(\xi))}\approx1-\sigma_{\Phi_{\mathrm{f}}}^{2}(1-\gamma(\xi))\\
 & \:\hat{\approx}\:\sigma_{\Phi_{\mathrm{f}}}^{2}\gamma(\xi)+(1-\sigma_{\Phi_{\mathrm{f}}}^{2})\:,
\end{align*}
i.e., the Yashiro-Lynch theory is qualitatively consistent with the
expectations from classic scattering theory, yet differs in detail.
First, SAXS theory is inherently founded on a single scattering assumption,
i.e., a thin sample assumption. This gives rise to the linear vs.\
exponential relation with the autocorrelation function. The constant
offset $(1-\sigma_{\Phi_{\mathrm{f}}}^{2})$ corresponds, within the
model employed by Strobl and Prade et al.\ \cite{Strobl2014,Prade2015},
to the unscattered fraction of the incident radiation. 

\subsection{Lambert-Beer relation for tomographic imaging}

With respect to tomographic imaging, linearity of the given contrast
modality is a necessary prerequisite. I.e., the signal generated by
a stack of samples must equal the sum of their isolated signals. Analogously,
the signal response is required to be proportional to the thickness
of a homogeneous sample. In classic X-ray computed tomography, this
is provided by the well known Lambert-Beer law of intensity attenuation.

An equivalent relation is also found for the darkfield contrast modality:
most directly, it can be inferred from the Gaussian convolution model
given previously in Section~\ref{subsec:DF-Interpretation}. Under
the assumption that multiple samples (sections of a larger sample),
characterized by their effective point spread widths $\sigma_{\phi,i}$,
don't interact, their combined effect will be a series of convolutions
(as direct consequence of their individual effects). As convolutions
of Gaussian kernels are additive in their variances, the combined
signal is then given by:
\begin{equation}
v=\prod_{i}e^{-\frac{\sigma_{\phi,i}^{2}}{2}}=e^{-\frac{1}{2}\sum_{i}\sigma_{\phi,i}^{2}}\:.
\end{equation}
This argument has been employed e.g.\ by Wang et al.\ \cite{Wang2009}
(in analogy to Khelashvili et al\@.\ \cite{Khelashvili2006}) and
Bech et al.\ \cite{Bech2010} to derive the feasibility of darkfield
tomography. The relation can further be generalized to non-Gaussian
convolution kernels by means of the convolution theorem, as has been
done by Modregger et al.\ \cite{Modregger2014PRL,Modregger2017PRL}:
\begin{align}
g(\Delta\phi) & =g_{1}(\Delta\phi)\circledast g_{2}(\Delta\phi)\circledast\cdots\nonumber \\
v'(q)=\mathcal{F}(g(\Delta\phi)) & =\prod_{\smash[b]{i}}\mathcal{F}(g_{i}(\Delta\phi))\\
 & =\smash[t]{\exp\Bigl(\sum_{i}\ln\mathcal{F}(g_{i})\Bigr)}\nonumber 
\end{align}
with $\circledast$ denoting the convolution operation and $\mathcal{F}$
the Fourier transformation. $q$ is the conjugate variable to $\Delta\phi$
after transformation. $g(\Delta\phi)$ is the effective point spread
kernel found for a sample, with $g_{i}(\Delta\phi)$ being individual
contributions. The prime on $v'(q)$ indicates the missing normalization
to $v'(0)$. The Gaussian convolution model is in fact a special case
of this more general formulation.  The frequency ($q$) dependence
of $v$ is usually not explicitly addressed in the majority of articles
on the subject, as $v$ is commonly explicitly defined as the visibility
of the first harmonic of the grating period, i.e., $v=v'(q_{1})/v'(q_{0})$
(cf.\ Pfeiffer et al.\ (2008) \cite{Pfeiffer2008}).

The linear superposition of exponential arguments is, independent
of the above considerations, also found by Yashiro et al.\ \cite{Yashiro2010}
and Lynch et al\@.\ \cite{Lynch2011} within their wave optical
derivations of darkfield contrast origination (cf.\ Section~\ref{subsec:DF-abinitio}).

\section{Anisotropic Darkfield Contrast\label{sec:aniso-df}}

The darkfield contrast is an oriented effect. As has been described
in the previous sections, it arises from refractive effects below
the spatial resolution of the imaging system and can be understood
in terms of the scattering cross section and autocorrelation function
of the sample's substructure. So far, the actual direction of autocorrelation
was not explicitly discussed and rather implicitly determined by the
orientation of the interferometer gratings. In fact, the scalar visibility
contrast found in a particular experiment is to be understood as a
feature of a higher dimensional quantity, which may or may not be
isotropic. For samples exhibiting structural anisotropy in the plane
perpendicular to the optical axis (i.e., parallel to the interferometer
gratings), the darkfield signal will vary when rotating the sample
(or the interferometer) about the optical axis due to the variation
in characteristic length scales along the instrument's direction of
sensitivity.

This anisotropy has initially been experimentally demonstrated by
Wen et al.\ \cite{HWen2009}, Jensen et al\@.\ \cite{Jensen2010PMB,Jensen2010PRB},
Revol et al.\ \cite{Revol2012}, Potdevin et al\@.\ \cite{Potdevin2012}
and Schaff et al\@.\ \cite{Schaff2014}, who considered planar samples
exhibiting highly ordered fibrous structures perpendicular to the
optical axis, such as carbon fibers, wood fibers, dentinal tubules
or trabecular bone. Technically, these cases all address distinctive
variations of the autocorrelation width of long fibers with respect
to their orientation. Yashiro et al.\ \cite{Yashiro2011} considered
the case of moderately anisotropic structures within an extended sample
and investigated the individual effects of autocorrelation width,
scattering cross section and autocorrelation shape (by both rotating
the sample and sampling multiple correlation lengths, cf.\ Sections
\ref{subsec:DF-abinitio} and \ref{subsec:DF-SAXS}). A systematic
experimental observation of darkfield signal variation in dependence
of the 3D orientation -- i.e., also considering inclinations with
respect to the optical axis -- of highly oriented fibers was presented
by Bayer et al.\ \cite{Bayer2012OE}.

The theoretic modeling of these anisotropy effects has been treated
phenomenologically using sinusoids reproducing the periodicity, phase
and an assessment of the degree of anisotropy of the signal in dependence
of a single orientation angle. This representation straight forwardly
enables planar vector radiographs, augmenting 2D X-ray images with
directional information on the unresolved substructure in the detection
plane. Mathematically, it is equivalent to a linear approximation
of the directional dependencies (as has also been pointed out by Jensen
et al.\ \cite{Jensen2010PRB}). Malecki et al.\ \cite{Malecki2013}
further investigated the sinusoid approximation in numerical simulations
of fibrous structures parallel to the detection plane. Current approaches
extending the concept of directional imaging to tomography on volumetric
samples (cf.\ \cite{Bayer2014,Malecki2014,Vogel2015,Wieczorek2016,Dittmann2017,Graetz2019})
derive from these planar signal models and thus implicitly presume
that the autocorrelation width perpendicular to the optical axis is
the principal origin of orientation dependency of the darkfield signal.
A first heuristic model of darkfield anisotropy extending beyond autocorrelation
effects and also accounting for variations in scattering cross section
has recently been presented parallel to the present work by Felsner
et al.\ \cite{Felsner2020}.

In the following, a model of general darkfield anisotropy for arbitrarily
oriented scatterers shall be derived based on the wave optical considerations
given by Yashiro and Lynch, thereby implicitly accounting both for
autocorrelation and scattering cross section dependencies.

\subsection{Anisotropic Gaussian density model of anisotropic scatterers\label{subsec:gaussian-ellipsoid-model}}

In preparation to advances in X-ray darkfield tensor tomography, the
anisotropy properties with respect to general sample orientations
shall be reviewed based on darkfield signal origination discussed
previously in Section~\ref{subsec:DF-abinitio}. The anisotropy of
arbitrary elongated structures shall be modeled by means of an anisotropic
Gaussian mass (or electron) density distribution
\begin{equation}
\rho(\vec{r})=\rho_{0}e^{-\frac{1}{2}\vec{r}\boldsymbol{T}\vec{r}}
\end{equation}
with $\vec{r}$ denoting a point in three dimensional space and 
\begin{equation}
\boldsymbol{T}=\left[\begin{array}{ccc}
T_{\mathrm{xx}} & T_{\mathrm{xy}} & T_{\mathrm{xz}}\\
T_{\mathrm{xy}} & T_{\mathrm{yy}} & T_{\mathrm{yz}}\\
T_{\mathrm{xz}} & T_{\mathrm{yz}} & T_{\mathrm{zz}}
\end{array}\right]=\boldsymbol{R}\left[\begin{array}{ccc}
\sigma_{1}^{-2} & 0 & 0\\
0 & \sigma_{2}^{-2} & 0\\
0 & 0 & \sigma_{3}^{-2}
\end{array}\right]\boldsymbol{R}^{T}\label{eq:density-tensor}
\end{equation}
being a symmetric, positive definite tensor characterized by positive
eigenvalues ($\sigma_{1}^{-2},\sigma_{2}^{-2},\sigma_{3}^{-2}$, representing
the inverse variances of $\rho(\vec{r})$) and a unitary rotation
matrix $\boldsymbol{R}$ (with the supscript $T$ denoting the transpose
operation). $\boldsymbol{R}$ defines the orientation of the Gaussian
ellipsoid object characterized by three orthogonal standard deviations
$\sigma_{1}$, $\sigma_{2}$ and $\sigma_{3}$. The Gaussian mass
density $\rho(\vec{r})$ may be interpreted as description of a stochastic
ensemble of smaller structures, or, more abstractly, as a generating
structure to approximate the phase shifting properties along the optical
axis (Eqs.~\ref{eq:integral-phaseshift}--\ref{eq:phase-T}) and
resulting autocorrelation function (Eq.~\ref{eq:aniso-auto-correlation})
of convex anisotropic structures irrespective of the precise microscopic
material distribution (e.g., a cylindrical fiber). It may in this
respect also be interpreted as a generalization of the Gaussian approximation
of the autocorrelation properties of spherical objects (cf.\ Section~\ref{subsec:DF-diffusion-approx}).

\subsection{Scattering cross section\label{subsec:ansio-DF-crosssection}}

\begin{figure}
\begin{centering}
\includegraphics[width=0.32\textwidth]{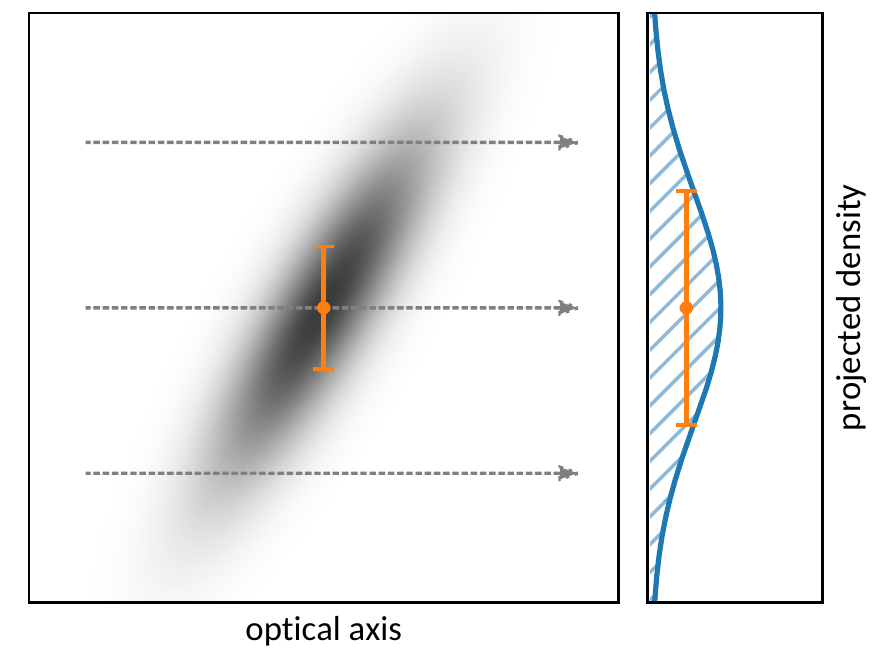}~\includegraphics[width=0.32\textwidth]{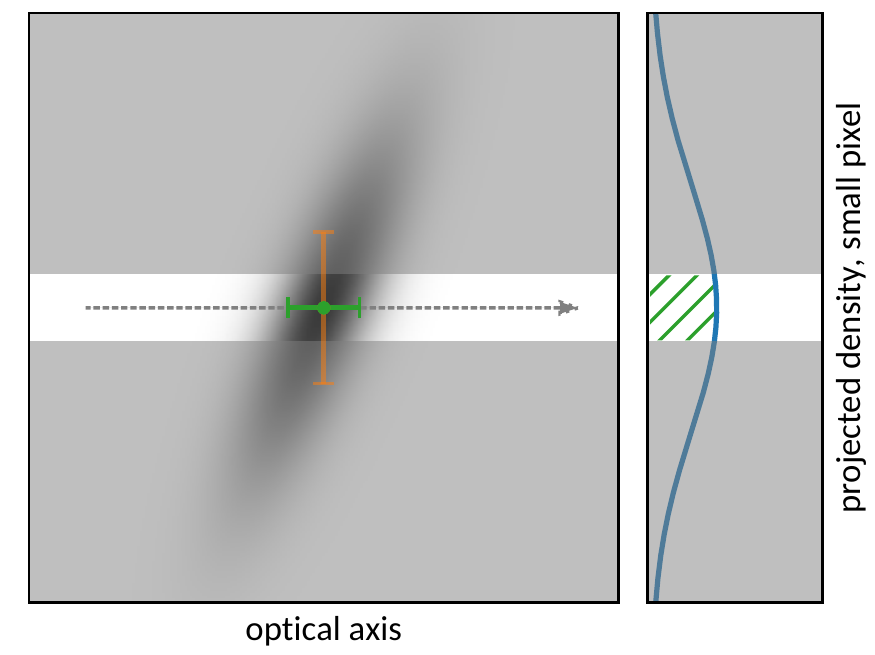}~\includegraphics[width=0.352\textwidth]{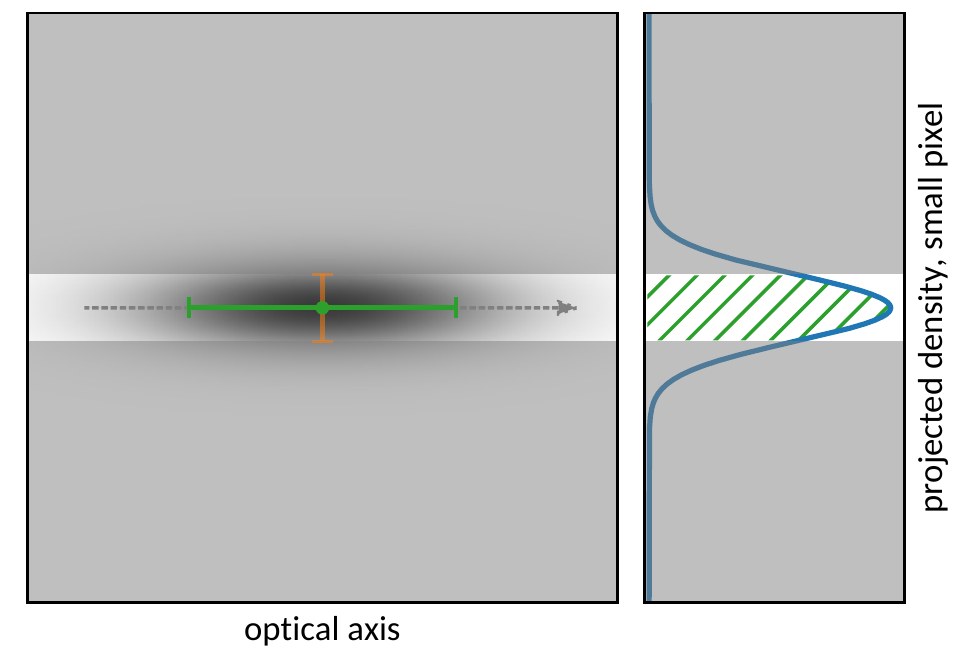}
\par\end{centering}
\caption{\label{fig:DF-scattering-cross-section}Illustration of the projection
of a Gaussian mass density distribution along the optical axis (\emph{left}).
The standard deviation of the projection will typically be larger
than the distribution's standard deviation parallel to the projection
plane (both marked in orange for comparison, cf.\ Eqs.~\ref{eq:integral-phaseshift}--\ref{eq:phase-T}).
When considering the case of pixels smaller than an elongated object
(with regions outside that pixel shaded in gray), the projected volume
contributing to an individual pixel will be confined by the pixel
itself (\emph{center and right}). Variations in the total projection
(hatched) affecting the scattering cross section captured by that
pixel are then dominated by the extent of the mass distribution along
the optical axis (indicated in green). Cf.\ Eq.~\ref{eq:small-obj-scatter-cross-section}
vs.\ Eq.~\ref{eq:large-object-scatter-cross-section}.}
\end{figure}
The scattering cross section, as has been discussed in Section~\ref{subsec:DF-abinitio},
is governed by the variance of phase front fluctuations $\sigma_{\Phi_{\mathrm{f}}}^{2}$
below the scale of the imaging system's spatial resolution. The total
phase shift $\Phi(x,y)$ along the optical axis (here: $z$) caused
by the presumed Gaussian density distribution is given by:
\begin{align}
\Phi(x,y) & =\Phi_{0}\int_{-\infty}^{\infty}e^{-\frac{1}{2}\vec{r}\boldsymbol{T}\vec{r}}\mathrm{d}z\nonumber \\
 & =\Phi_{0}\sqrt{\frac{2\pi}{T_{\mathrm{zz}}}}e^{-\frac{1}{2}\vec{r}\boldsymbol{T}_{\Phi}\vec{r}}\label{eq:integral-phaseshift}
\end{align}
with 
\begin{equation}
\boldsymbol{T}_{\Phi}=\left[\begin{array}{ccc}
T_{\mathrm{xx}}-\frac{T_{\mathrm{xz}}^{2}}{T_{\mathrm{zz}}} & T_{\mathrm{xy}}-\frac{T_{\mathrm{xz}}T_{\mathrm{yz}}}{T_{\mathrm{zz}}} & 0\\
T_{\mathrm{xy}}-\frac{T_{\mathrm{xz}}T_{\mathrm{yz}}}{T_{\mathrm{zz}}} & T_{\mathrm{yy}}-\frac{T_{\mathrm{yz}}^{2}}{T_{\mathrm{zz}}} & 0\\
0 & 0 & 0
\end{array}\right]\label{eq:phase-T}
\end{equation}
and $\Phi_{0}$ being a constant factor resulting from the Gaussian
density distribution's maximum value $\rho_{0}$ and the material's
refractive index for the considered wavelength. The difference between
$T_{xx}$ and $T_{xx}-\frac{T_{xz}^{2}}{T_{zz}}$ is illustrated in
Figure~\ref{fig:DF-scattering-cross-section} (left).

When defining the local mean variance used by Yashiro and Lynch et
al.\ \cite{Yashiro2010,Lynch2011} within the discussion of darkfield
origination (cf.\ Section \ref{subsec:DF-abinitio})
\begin{equation}
\sigma_{\Phi_{\mathrm{f}}}^{2}=\overline{(\Phi_{\mathrm{f}}(x,y)-\overline{\Phi_{\mathrm{f}}(x,y)})^{2}}
\end{equation}
 by means of a Gaussian weighting kernel accounting for the imaging
point spread function's width $\sigma_{\mathrm{PSF}}$ and assuming
this point spread to be larger than the considered structure sizes
$\sigma_{i}$, i.e., $\sigma_{\mathrm{PSF}}/\sigma_{i}\gg1$, we find:
\begin{align}
\sigma_{\Phi_{\mathrm{f}}}^{2} & \approx\frac{1}{2\pi\sigma_{\mathrm{PSF}}}\int_{-\infty}^{\infty}e^{-\frac{1}{2}\frac{x^{2}+y^{2}}{\sigma_{\mathrm{PSF}}^{2}}}\left(\Phi(x,y)-\frac{1}{2\pi\sigma_{\mathrm{PSF}}}\int_{-\infty}^{\infty}e^{-\frac{1}{2}\frac{x^{2}+y^{2}}{\sigma_{\mathrm{PSF}}^{2}}}\Phi(x,y)\:\mathrm{d}x\,\mathrm{d}y\right)^{2}\mathrm{d}x\,\mathrm{d}y\\
 & \approx\frac{1}{2\pi\sigma_{\mathrm{PSF}}}\int_{-\infty}^{\infty}e^{-\frac{1}{2}\frac{x^{2}+y^{2}}{\sigma_{\mathrm{PSF}}^{2}}}\left(\Phi(x,y)-\frac{1}{2\pi\sigma_{\mathrm{PSF}}}\int_{-\infty}^{\infty}\Phi(x,y)\:\mathrm{d}x\,\mathrm{d}y\right)^{2}\mathrm{d}x\,\mathrm{d}y\nonumber \\
 & \approx\frac{\Phi_{0}\pi}{\sigma_{\mathrm{PSF}}\sqrt{\det(\boldsymbol{T})}}\left(\frac{1}{\sigma_{\mathrm{PSF}}\sqrt{T_{\mathrm{zz}}}}-\frac{2}{\sigma_{\mathrm{PSF}}^{3}\sqrt{\det(\boldsymbol{T})}}\right)\nonumber \\
 & \approx\frac{\Phi_{0}\pi}{\sigma_{\mathrm{PSF}}^{2}\sqrt{\det(\boldsymbol{T})}}\frac{1}{\sqrt{T_{\mathrm{zz}}}}\:,
\end{align}
and can thus finally state:\vspace{-0.75em}
\begin{align}
\sigma_{\Phi_{\mathrm{f}}}^{2} & \approx\frac{1}{\sqrt{\hat{n}\boldsymbol{T}\hat{n}}}\overbrace{\left(\frac{\Phi_{0}\pi}{\sigma_{\mathrm{PSF}}^{2}\sqrt{\det(\boldsymbol{T})}}\right)}^{\mathclap{\text{orientation independent}}}\quad\text{for objects smaller than the pixel size}.\label{eq:small-obj-scatter-cross-section}
\end{align}
$\sqrt{T_{zz}}^{-1}$ was identified here as the standard deviation
along the optical axis and is therefore in more general terms described
by $\sqrt{\hat{n}\boldsymbol{T}\hat{n}}^{-1}$ for arbitrary orientations
$\hat{n}$ of the optical axis ($\bigl\Vert\hat{n}\bigr\Vert=1$).
The determinant $\det(\boldsymbol{T})=(\sigma_{1}\sigma_{2}\sigma_{3})^{-2}$
is invariant under rotations $\boldsymbol{R}$ and therefore, as $\Phi_{0}$,
an invariant factor with respect to the orientation dependence of
$\sigma_{\Phi_{\mathrm{f}}}^{2}$ (provided the density distribution
fits into the considered integration range defined by $\sigma_{\mathrm{PSF}}$).

The following conclusions can be drawn from these results on $\sigma_{\Phi_{\mathrm{f}}}^{2}$.
It is proportional to the extent $\sqrt{\hat{n}\boldsymbol{T}\hat{n}}^{-1}$
of the considered structure along the optical axis, and with $\sqrt{\det(\boldsymbol{T})}^{-1}=\sigma_{1}\sigma_{2}\sigma_{3}$
further proportional to its volume.  In the eigenbasis of $\boldsymbol{T}$,
the relation $\sigma_{\Phi_{\mathrm{f}}}^{2}\propto\sqrt{\det(\boldsymbol{T})\,\hat{n}\boldsymbol{T}\hat{n}}^{-1}$
simplifies to $\sigma_{\Phi_{\mathrm{f}}}^{2}\propto(\sigma_{1}\sigma_{2})\sigma_{3}^{2}$:
the variance $\sigma_{\Phi_{\mathrm{f}}}^{2}$ is directly proportional
to the variance $\sigma_{3}^{2}$ of the underlying density distribution
along the optical axis and scales with its cross sectional area, which
is proportional to $\sigma_{1}\sigma_{2}$.

In the limit of objects extending way beyond the size of a detector
pixel (e.g., long fibers), the cross sectional area of the object's
projection onto a pixel becomes almost independent of its orientation,
as the considered area is rather confined by the detection area of
the pixel itself, which is proportional to $\sigma_{\mathrm{PSF}}^{2}$
(cf.\ Figure~\ref{fig:DF-scattering-cross-section}). In these cases,
the orientation dependence of the scattering cross section affecting
that pixel is thus, given the previous observations, dominated by
the variance of the density distribution along the optical axis, while
the cross sectional area is bounded by $\sigma_{\mathrm{PSF}}^{2}$,
as opposed to the extent of the material distribution:
\begin{equation}
\sigma_{\Phi_{\mathrm{f}}}^{2}\propto\frac{1}{\hat{n}\boldsymbol{T}\hat{n}}\overbrace{\left(\frac{\Phi_{0}\pi}{\sigma_{\mathrm{PSF}}^{2}\sigma_{\mathrm{PSF}}^{2}}\right)}^{\mathclap{\text{orientation independent}}}\quad\text{for objects larger than the pixel size}.\label{eq:large-object-scatter-cross-section}
\end{equation}

\subsection{Autocorrelation function\label{subsec:aniso-DF-autocorrelation}}

The normalized planar autocorrelation $\gamma_{\mathrm{2D}}(\vec{\xi})$
corresponding to $\Phi(x,y)$ is, based on the additivity of variances
for convolutions of Gaussians, given by
\begin{equation}
\gamma_{\mathrm{2D}}(\vec{\xi})=e^{-\frac{1}{4}\vec{\xi}\boldsymbol{T}_{\Phi}\vec{\xi}}\:,
\end{equation}
assuming $\vec{\xi}$ to lie in the $x$-$y$ plane perpendicular
to the optical axis ($z$). When assuming that all relative rotations
of interferometer and object are accounted for in $\boldsymbol{R}$
and thus in $\boldsymbol{T}$ and $\boldsymbol{T}_{\Phi}$, the autocorrelation
direction may without loss of generality be defined parallel to the
$x$ axis, i.e., $\vec{\xi}=(\xi,0,0)$, such that 
\begin{equation}
\gamma(\xi)=e^{-\frac{1}{4}\xi^{2}(T_{\mathrm{xx}}-\frac{T_{\mathrm{xz}}^{2}}{T_{\mathrm{zz}}})}\:.\label{eq:aniso-auto-correlation}
\end{equation}

As the mass density distribution characterized by $\boldsymbol{T}$
by design already represents all structures within the optical path
relevant to the considered detector pixel, no further averaging of
autocorrelation properties over the pixel area is required. The autocorrelation
width moreover is an immanent property of the mass density distribution,
and thus in contrast to the scattering cross section not limited by
the pixel extent $\sigma_{\mathrm{PSF}}$.

For isotropic density distributions ($\sigma_{i}=\sigma$), the off-diagonals
of $\boldsymbol{T}$ and $\boldsymbol{T}_{\mathrm{\Phi}}$ will be
zero. The autocorrelation then reduces to $\gamma(\xi)=\exp(-\frac{1}{4}\frac{\xi^{2}}{\sigma^{2}})$,
corresponding directly to the 3D autocorrelation function of an isotropic
Gaussian density distribution without explicit averaging over the
optical axis. In order to provide further intuition to the modeled
Gaussian density distribution, this result may be compared to the
approximate autocorrelation function for spheres of diameter $D$
($\gamma(\xi)\approx\exp(-\frac{1}{2}\frac{\xi^{2}}{(D/3)^{2}})$,
cf.\ Eq.~\ref{eq:gaussian-sphere-corr-approx})). I.e., an isotropic
Gaussian mass density distribution with standard deviation $\sigma$
can be interpreted as modeling a sphere of diameter $D\approx4\sigma$.

\subsection{Model of darkfield contrast anisotropy\label{subsec:aniso-model}}

Substituting the above results (Eqs.\ \ref{eq:small-obj-scatter-cross-section}--\ref{eq:aniso-auto-correlation})
on scattering cross section and autocorrelation into Eq.~\ref{eq:yashiro-visibility},
the fringe contrast visibility is given by:
\begin{equation}
v_{\mathrm{lp}}\approx\exp\Biggl(-\frac{1}{\sqrt{\hat{n}\boldsymbol{T}\hat{n}}}\overbrace{\left(\frac{\Phi_{0}\pi}{\sigma_{\mathrm{PSF}}^{2}\sqrt{\det(\boldsymbol{T})}}\right)}^{\mathclap{\text{orientation independent}}}\left(1-e^{-\frac{1}{4}\vec{\xi}\,\boldsymbol{T}_{\Phi}\vec{\xi}\,}\right)\Biggr)
\end{equation}
which for scatterers, such as long fibers, larger than the typical
integration width (e.g., pixel size) characterized by $\sigma_{\mathrm{PSF}}$,
modifies to 
\begin{equation}
v_{\mathrm{sp}}\approx\exp\Biggl(-\frac{1}{\hat{n}\boldsymbol{T}\hat{n}}\overbrace{\left(\frac{\Phi_{0}\pi}{\sigma_{\mathrm{PSF}}^{4}}\right)}^{\mathclap{\text{orientation independent}}}\left(1-e^{-\frac{1}{4}\vec{\xi}\,\boldsymbol{T}_{\Phi}\vec{\xi}\:}\right)\Biggr)
\end{equation}
with $\boldsymbol{T}$ characterizing the inverse variances of the
considered scatterers' density distribution (Eq.~\ref{eq:density-tensor}),
$\hat{n}$ the unit vector along the optical path, $\vec{\xi}$ the
oriented autocorrelation distance and $\boldsymbol{T}_{\Phi}$ (Eq.~\ref{eq:phase-T})
describing the inverse variances of the scatterers' total phase shift
along the optical axis (Eq.~\ref{eq:integral-phaseshift}), based
on their density distribution characterized by $\boldsymbol{T}$.
The subscripts ``sp'' and ``lp'' indicate the ``small pixel''
and ``large pixel'' cases as compared to the extent of the considered
scatterers characterized by $\boldsymbol{T}$.

When defining, without loss of generality, $\hat{\boldsymbol{n}}$
parallel to the $z$-axis and $\vec{\xi}$ parallel to the $x$-axis
(assuming all rotations to be considered within $\boldsymbol{T}$,
cf.\ Eq.~\ref{eq:density-tensor}), and further using the first
order approximation $1-\exp(-u)\approx u$ for small $u$, then the
following simplified proportionality relations for the expected orientation
dependence of the darkfield contrast $\mu_{\mathrm{DF}}=-\ln(v)$
result for the large pixel (lp) and small pixel (sp) cases respectively:
\begin{align}
-\ln(v_{\mathrm{lp}}) & \appropto\frac{1}{\sqrt{T_{zz}}}(T_{\mathrm{xx}}-\frac{T_{\mathrm{xz}}^{2}}{T_{\mathrm{zz}}})\label{eq:aniso-model-lp}\\
-\ln(v_{\mathrm{sp}}) & \appropto\hphantom{\frac{1}{\sqrt{T_{zz}}}}\mathllap{\frac{1}{T_{zz}}\,}(T_{\mathrm{xx}}-\frac{T_{\mathrm{xz}}^{2}}{T_{\mathrm{zz}}})\:.\label{eq:aniso-model-sp}
\end{align}

\section{Experimental verification}

With respect to applications in quantitative darkfield tomography,
the Yashiro-Lynch results are explicitly reproduced in a tomography
setting using a phantom containing spherules of various diameters.
Furthermore, the derived anisotropy properties are reproduced using
a carbon fiber phantom, with particular focus on the role of the scattering
cross section in addition to the autocorrelation width. The experiments
have been performed at the ID19 beamline of the European Synchrotron
Radiation Facility (ESRF) using the Talbot-Interferometer by Weitkamp
et al.\ \cite{Weitkamp2010}. A $\pi$-shifting phase grating G$_{1}$
of 4.8\textmu m period and an absorbing analyzer grating G$_{2}$
of 2.4\textmu m period were used in a monochromatic 35keV setting.
Images were sampled at an effective pixel size of 61.6\textmu m.

\subsection{Feature size and positional dependence of darkfield contrast\label{subsec:DF-experiment-distance-dependence}}

\begin{figure}
\centering{}\includegraphics[width=0.48\textwidth]{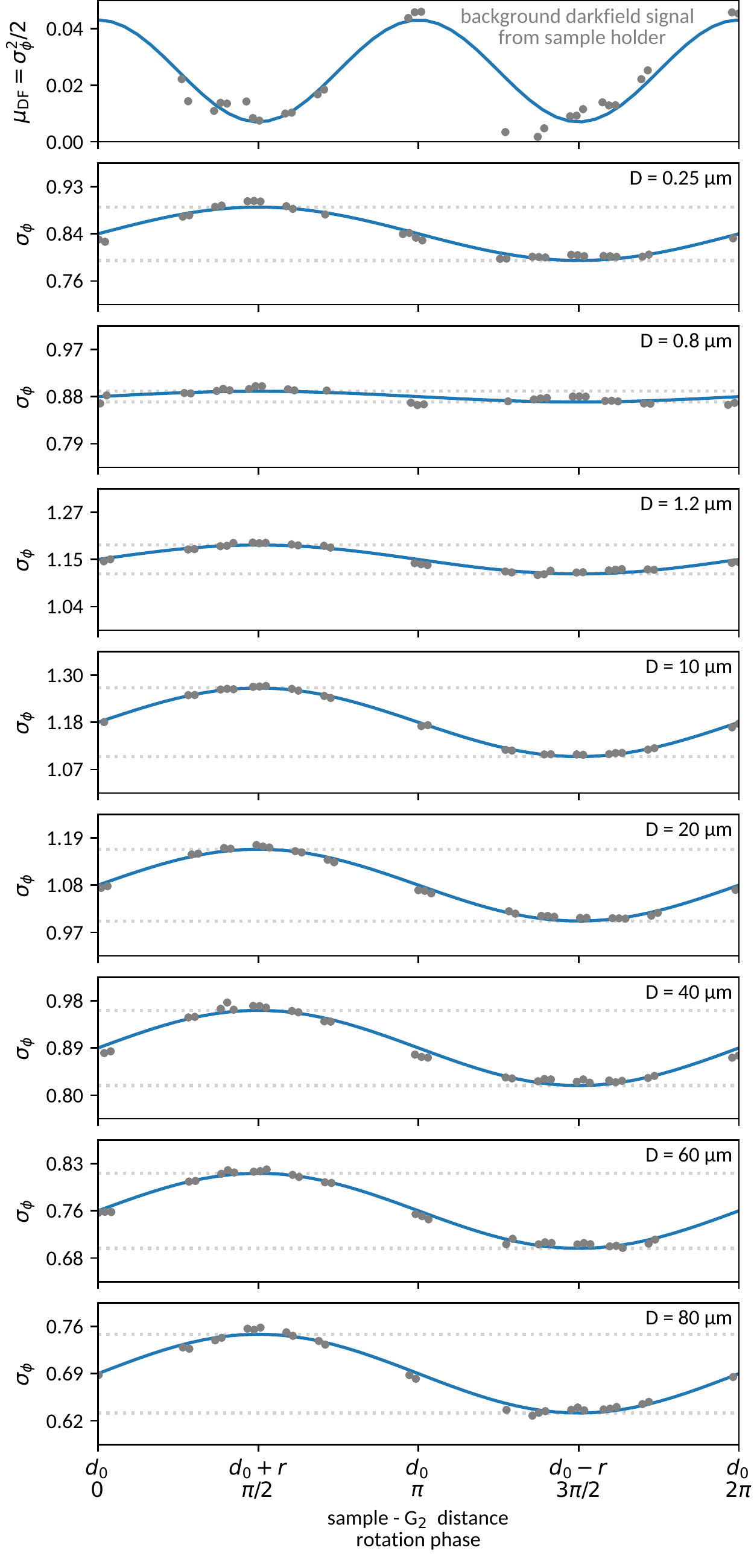}~~~~~~%
\begin{minipage}[b][1\totalheight][t]{0.5\columnwidth}%
\begin{center}
~~~\includegraphics[width=0.87\textwidth]{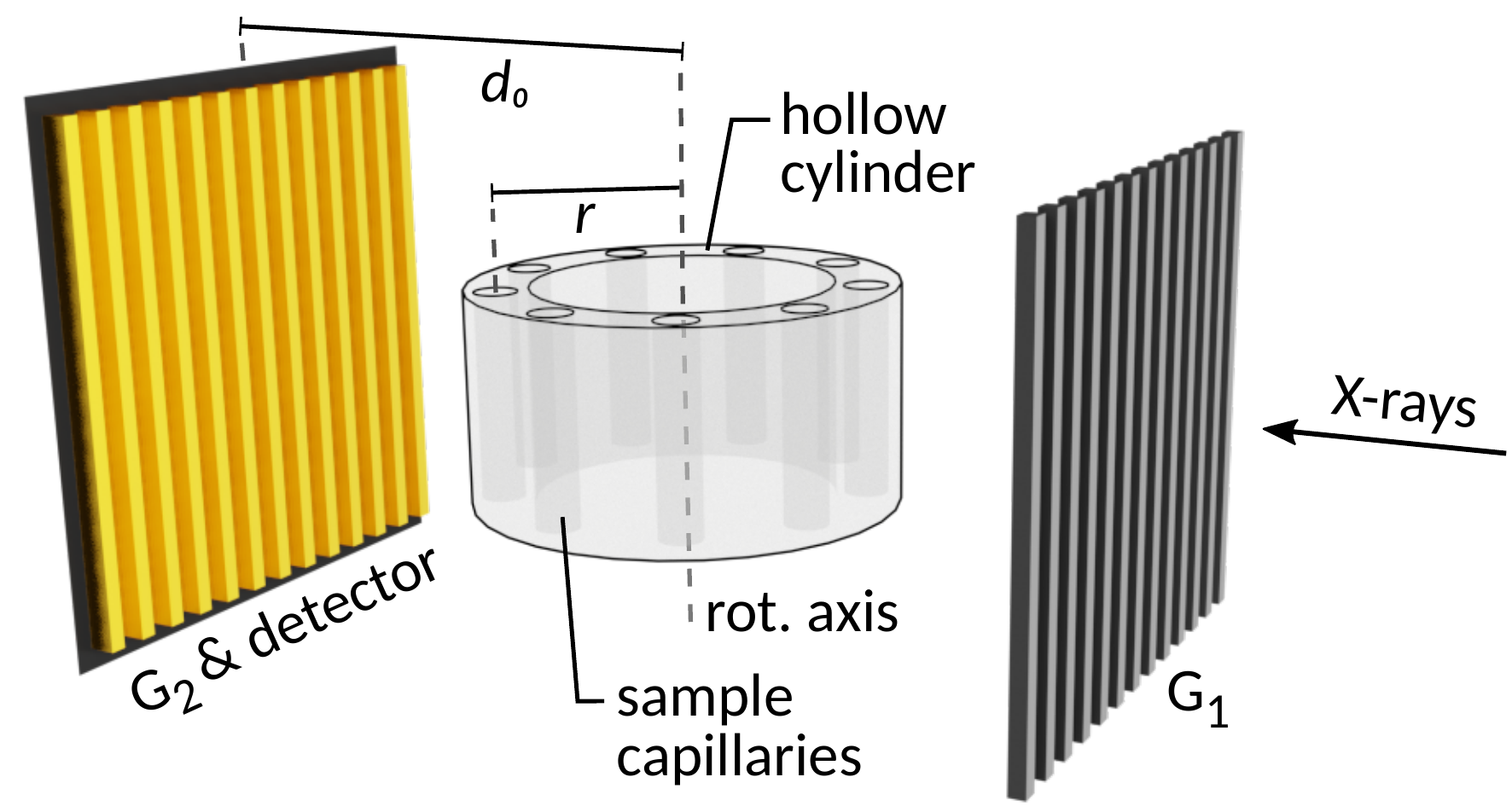}
\par\end{center}
\begin{center}
~~~\includegraphics[viewport=0bp 0bp 288bp 165bp,width=0.88\textwidth]{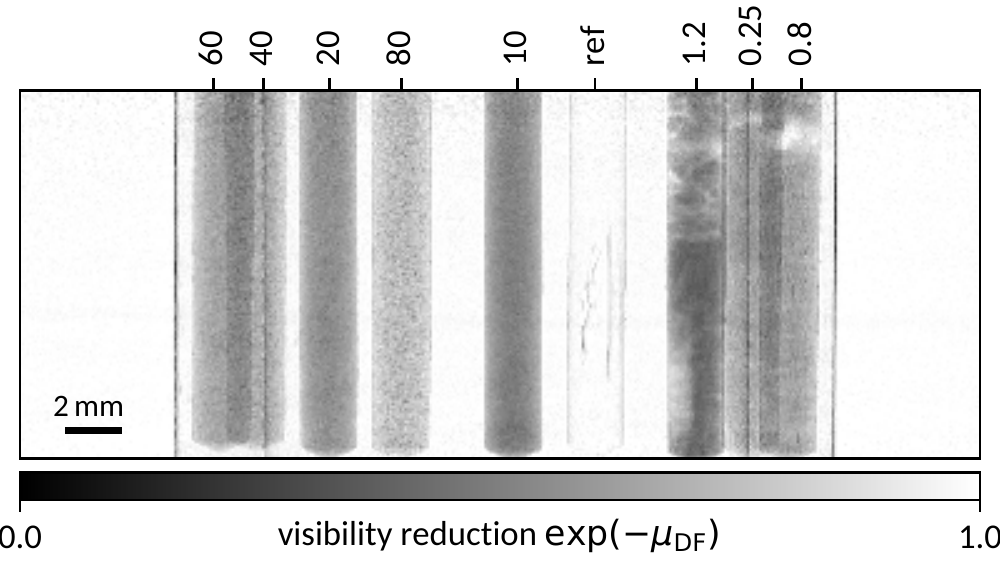}
\par\end{center}
\begin{center}
\includegraphics[width=0.9\textwidth]{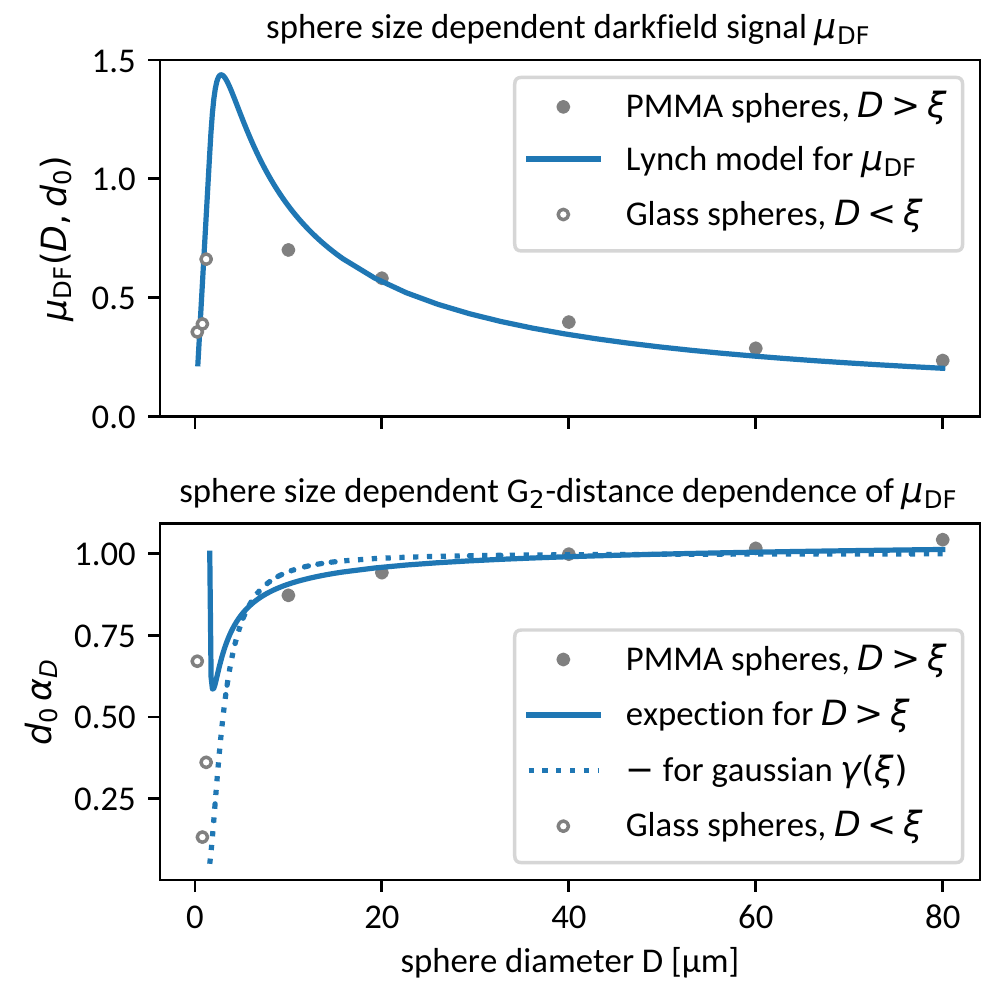}\\
~
\par\end{center}%
\end{minipage}\caption{\label{fig:DF-distance-dependence}Experimental data (left) on the
dependence of darkfield contrast both on the distance $d$ between
sample and analyzer grating G$_{2}$ and on the structure size. Sinusoidal
distance variations for multiple samples were realized by rotating
a cylindrical container of radius $r\approx{}$1$\,$cm comprising
nine capillary tubes arranged equidistantly along the perimeter, eight
of which are filled with spherules of diameters $D$ ranging between
0.25$\,$μm to 80$\,$μm (cf.\ sketch on the upper right). The ninth
empty one serves as reference for the background signal generated
by the sample container, which is approximated using a $\cos^{2}$
function due to the expected 180° symmetry. Darkfield signals are
evaluated for all viewing angles which allow for unobstructed views
on individual capillaries (hence the uneven sampling pattern), and
the background signal is subtracted prior to further processing. An
example projection image is shown on the right. As the distance dependence
is expected to be linearly approximable for $\sigma_{\phi}=\sqrt{2\mu_{\mathrm{DF}}}$
(cf.\ Eqs.\ \ref{eq:DF-v-exp-sigma2} and \ref{eq:yashiro-sigma-approx}),
the signals are evaluated accordingly and approximated by a sinusoid
corresponding to the variation in sample-G$_{\mathrm{2}}$ distance.
Quantitative comparisons of both the spherule size dependence and
the distance dependence of the darkfield signal with the respective
models given  in Eqs\@.\ \ref{eq:lynch-darkfield-function}, \ref{eq:alpha_D-relation}
(with Eq.~\ref{eq:yashiro-correlation-function}) and \ref{eq:alpha_D-relation-gaussian}
are shown on the lower right. Deviations of $\mu_{\mathrm{DF}}(D,d_{0})$
from the expected model (in particular at $D={}$10$\,$\textmu m)
can be related to unaccounted variations in the volume fraction of
spheres, cf.\ Figures \ref{fig:DF-tomography}--\ref{fig:DF-tomography-quantitative}.
The $d$-dependence in cases of $D<\xi$ (particularly pronounced
for $D={}$0.25$\,$\textmu m) can be attributed to the non-vanishing
autocorrelation function of the sphere arrangement rather than the
spheres themselves.}
\end{figure}

While the dependence of the system's correlation length $\xi$ on
the distance $d$ between sample and analyzer grating G$_{2}$ is
beneficial with respect to quantitative material characterization
analog to other scattering techniques, the resulting distance dependence
of the darkfield signal is an undesired perturbation in the context
of tomographic darkfield imaging. For centimeter scaled samples, the
sample extent is typically not negligible anymore with respect to
its mean distance $d_{0}$ from the analyzer grating, wherefore the
distance dependence is explicitly examined using a rotating circular
arrangement of differently sized spherules and compared to the theoretical
expectations discussed in Sections \ref{subsec:DF-diffusion-model}--\ref{subsec:DF-diffusion-approx}.
Figure~\ref{fig:DF-distance-dependence} shows a sketch of the experimental
setup and a summary of the respective results.

The distance and sphere diameter dependent signals (Fig.~\ref{fig:DF-distance-dependence}
left) extracted from projections of the rotating sample cylinder shall
be in particular compared to the explicit expressions for $\mu_{\mathrm{DF}}$
and the autocorrelation function $\gamma_{D}(\xi)$ for spherical
particles of diameter $D$ given by Lynch et al.\ \cite{Lynch2011}
and Yashiro et al.\ \cite{Yashiro2011} respectively:
\begin{align}
\mu_{\mathrm{DF}}(D,\xi) & =\Delta z\frac{3\pi^{2}}{\lambda^{2}}f\bigl|\Delta\chi\bigr|^{2}D\begin{cases}
\,1-\sqrt{1-\frac{\xi^{2}}{D^{2}}}\,(1+\frac{1}{2}\frac{\xi^{2}}{D^{2}})+(\frac{\xi^{2}}{D^{2}}-\frac{1}{4}\frac{\xi^{4}}{D^{4}}) & \text{for }D>\xi\\
\,1 & \text{for }D\le\xi
\end{cases}\label{eq:lynch-darkfield-function}\\
\gamma_{D}(\xi) & =(1+\frac{1}{2}\frac{\xi^{2}}{D^{2}})\sqrt{1-\frac{\xi^{2}}{D^{2}}}+(2\frac{\xi^{2}}{D^{2}}+\frac{1}{4}\frac{\xi^{4}}{D^{4}})\ln(\frac{\xi/D}{1+\sqrt{1-\xi^{2}/D^{2}}})\quad\text{for }D\ge\xi\label{eq:yashiro-correlation-function}\\
\xi & =\frac{\lambda}{T_{\mathrm{G}2}}d\quad(\text{cf.\ Eq.\ \ref{eq:correlation-length})}\,.\nonumber 
\end{align}
Constants of the material and experiment will be specified later.
Beforehand, the specific methodology of the present data analysis
shall be further outlined: By Eq.~\ref{eq:yashiro-sigma-approx},
the darkfield signal is expected to be approximately quadratic in
$d$. The data is thus, after subtraction of the background darkfield
signal caused by the sample container (cf.\ Fig.~\ref{fig:DF-distance-dependence},
upper left), considered in the square root domain, i.e.\ transformed
to $\sigma_{\phi}=\sqrt{2\mu_{\mathrm{DF}}}=\sqrt{-2\ln(v)}$. It
is then fitted to the first order Taylor expansion
\begin{align}
\sigma_{\phi}(D,d) & \approx\sigma_{\phi}(D,d_{0})\,(1+\alpha_{D}\underbrace{r\sin(\omega-\omega_{0})}_{\smash[b]{\Delta d(\omega)}})\label{eq:sigmaphi-sinusoid-model}\\
\intertext{\text{of the unknown actual function \ensuremath{\sigma_{\phi}(D,d)} about the mean sample-G\ensuremath{_{2}}-distance \ensuremath{d_{0}}, with}}\alpha_{D} & \,=\,\frac{1}{\sigma_{\phi}(D,d_{0})}\left.\frac{\partial}{\partial d}\sigma_{\phi}(D,d)\right|_{d=d_{0}}\:.\label{eq:sigmaphi-alpha-relation}
\end{align}
The parameters $r$ and $\omega-\omega_{0}$ denote the distance and
rotation phase of the considered sample capillary with respect to
the rotational axis, and $d_{0}+\Delta d(\omega)$ describes the resulting
orientation dependent sample-G$_{2}$-distance $d$. Deviations from
$\alpha_{D}=1/d_{0}$ correspond to deviations from the assumptions
of the linear diffusion model (Eqs.\ \ref{eq:DF-diffusion-sigma-theta},
\ref{eq:DF-linear-diffusion-coeff} and \ref{eq:yashiro-sigma-approx}),
as can be easily verified by comparison of Eq.~\ref{eq:sigmaphi-sinusoid-model}
with the Taylor expansion of the respective model $\sigma_{\phi}(D,d)\propto d\,\sigma_{\theta}(D)$. 

While the mean darkfield signal $\mu_{\mathrm{DF}}(D,d_{0})=\frac{1}{2}\sigma_{\phi}(D,d_{0})^{2}$
contained in Eq.~\ref{eq:sigmaphi-sinusoid-model} can now be directly
compared to Eq.~\ref{eq:lynch-darkfield-function} given an estimate
of the involved material constants (which will be provided later),
further transformations are required for a comparison of $\alpha_{D}$
with the $d(\omega)$-dependent autocorrelation function $\gamma_{D}(\xi(d))$
(Eq.~\ref{eq:yashiro-correlation-function}). By substituting the
relations
\begin{align*}
\sigma_{\phi} & =\sqrt{2\mu_{\mathrm{DF}}}=\sqrt{2\,\sigma_{\Phi_{\mathrm{f}}}^{2}(D)\,(1-\gamma(\xi(d)))}\\
\text{and}\quad\frac{\partial}{\partial d}\gamma(\xi(d)) & =\frac{\partial}{\partial\xi}\gamma(\xi)\frac{\partial\xi}{\partial d}=\frac{\partial}{\partial\xi}\gamma(\xi)\frac{\xi}{d}
\end{align*}
into Eq.~\ref{eq:sigmaphi-alpha-relation}, the following correspondence
results at $d=d_{0}$:
\begin{align}
\alpha_{D} & =-\frac{1}{2}\frac{\xi_{0}}{d_{0}}\frac{\frac{\partial}{\partial\xi}\gamma(\xi)|_{\xi=\xi_{0}}}{1-\gamma(\xi_{0})}\label{eq:alpha_D-relation}\\
\text{with}\quad\xi_{0} & =\frac{\lambda}{T_{\mathrm{G}2}}d_{0}\:,
\end{align}
which can be compared to the observed values of $\alpha_{D}$ determined
using Eq.~\ref{eq:sigmaphi-sinusoid-model}. The above expression
may either be evaluated using Eq.~\ref{eq:yashiro-correlation-function},
or may further be approximated using the Gaussian autocorrelation
model given in Eqs.~\ref{eq:gaussian-sphere-corr-approx}--\ref{eq:yashiro-sigma-approx}.
Eq.~\ref{eq:alpha_D-relation} then simplifies to
\begin{align}
\hphantom{\text{with }\xi_{0}}\mathllap{\alpha_{D}} & \approx\mathrlap{\frac{1}{2}\frac{1}{d_{0}}\frac{\xi_{0}^{2}}{(D/3)^{2}}\left(e^{\frac{1}{2}\frac{\xi_{0}^{2}}{(D/3)^{2}}}-1\right)^{-1}.}\hphantom{-\frac{1}{2}\frac{\xi_{0}}{d_{0}}\frac{\frac{\partial}{\partial\xi}\gamma(\xi)|_{\xi=\xi_{0}}}{1-\gamma(\xi_{0})}}\label{eq:alpha_D-relation-gaussian}
\end{align}
In the limit of large spheres, i.e., $D\to\infty$, the following
limit as expected by the linear diffusion model is approached both
for Eq.~\ref{eq:alpha_D-relation} with Eq. \ref{eq:yashiro-correlation-function}
as well as its approximation in Eq.~\ref{eq:alpha_D-relation-gaussian}:
\begin{equation}
\hphantom{\text{with }\xi_{0}}\mathllap{\lim_{\mathclap{D\rightarrow\infty}}\,\alpha_{D}}=\mathrlap{\frac{1}{d_{0}}\,.}\hphantom{-\frac{1}{2}\frac{\xi_{0}}{d_{0}}\frac{\frac{\partial}{\partial\xi}\gamma(\xi)|_{\xi=\xi_{0}}}{1-\gamma(\xi_{0})}}\label{eq:alpha_D-maximum}
\end{equation}

A comparison of the experimental data (Fig.~\ref{fig:DF-distance-dependence}
left) with Eqs.\ \ref{eq:lynch-darkfield-function}, \ref{eq:alpha_D-relation}
(using Eq.~\ref{eq:yashiro-correlation-function}) and \ref{eq:alpha_D-relation-gaussian}
is shown on the bottom right of Fig.~\ref{fig:DF-distance-dependence}.
In contrast to Eq.~\ref{eq:lynch-darkfield-function}, Eqs.\ \ref{eq:alpha_D-relation}--\ref{eq:alpha_D-relation-gaussian}
do have only one single free parameter, $d_{0}$. A least squares
fit yields 
\begin{align}
d_{0} & \approx\mathrlap{10.7\,\mathrm{cm}}\hphantom{-\frac{1}{2}\frac{\xi_{0}}{d_{0}}\frac{\frac{\partial}{\partial\xi}\gamma(\xi)|_{\xi=\xi_{0}}}{1-\gamma(\xi_{0})}}\label{eq:DF-distance-dependence-d0}\\
\hphantom{\text{with }\xi_{0}}\mathllap{\text{and thus }\xi_{0}} & \approx1.5\,\mathrm{\text{\textmu}m}
\end{align}
which is in excellent agreement with the actual experimental configuration
of $d_{0}=(0.1\pm0.01)\,\mathrm{m}$.

Note that the autocorrelation of a given finite shape (as the spheres)
inherently becomes zero for correlation distances larger than the
extent (diameter) of that shape, wherefore the submicrometer sized
glass spherules are not described by Eqs.\ \ref{eq:yashiro-correlation-function}
and \ref{eq:alpha_D-relation}. That they nevertheless exhibit a distance
dependence (cf.\ Fig.~\ref{fig:DF-distance-dependence}) can rather
be attributed to the fact that the dense packing of spheres of $0.25$
to $1.2\mathrm{\,\text{\textmu}m}$ diameter implies that the autocorrelation
of the sphere arrangement, i.e., the structure factor (see also Section~\ref{subsec:DF-SAXS}
and reference \cite{Gkoumas2016}), becomes non-negligible at the
given mean correlation distance $\xi_{0}$ of $1.5\,\mathrm{\text{\textmu}m}$.

Finally, with respect to the quantitative comparison of the actual
darkfield signals $\mu_{\mathrm{DF}}(D)$ at $d_{0}$ to the theoretic
prediction as given by Eq.~\ref{eq:lynch-darkfield-function}, the
following constants are used:
\begin{align*}
\lambda & =35.54\times10^{-12}\,\mathrm{m} & f & \approx0.74\\
T_{\mathrm{G}2} & =2.4\,\mathrm{\text{\textmu}m} & \Delta z & \approx10^{-3}\,\mathrm{m}\\
d=d_{0} & \approx10^{-1}\,\mathrm{m} & r_{e}\rho_{e,\mathrm{PMMA}} & \approx10^{15}\,\mathrm{m}^{-2}\\
\Delta\chi & \approx-\frac{\lambda^{2}}{2\pi}r_{e}\rho_{e} & \mathllap{\text{such that}}\;\,\Delta z\frac{3\pi^{2}}{\lambda^{2}}f\bigl|\Delta\chi\bigr|^{2} & \approx0.7\,\mathrm{\text{\textmu}m}^{-1}\,,
\end{align*}
with $\lambda$ being the X-ray wavelength (corresponding to 35keV),
$T_{\mathrm{G2}}$ the analyzer grating period, $d$ the sample-G$_{2}$
distance, $f$ the volume fill factor (assuming dense sphere packing),
$\Delta z$ the average sample thickness and $r_{e}\rho_{e,\mathrm{PMMA}}$
the scattering length density for PMMA.\footnote{estimated for C$_{5}$H$_{8}$O$_{2}$ with density of 1.2g/cm$^{3}$
from information provided by NIST for Cu and Mo K$_{\alpha}$ lines
at \texttt{https://www.ncnr.nist.gov/resources/activation/}} $\Delta\chi$ is the refractive index, whose imaginary part is, in
the present case, negligible with respect to its absolute magnitude.
Note the difference in notation with respect to \cite{Lynch2011}
regarding $\mu_{\mathrm{DF}}$ which here for consistency refers to
the actual darkfield signal as opposed to a thickness normalized darkfield
coefficient. Thickness is here explicitly accounted for by $\Delta z$. 

\subsection{Quantitative darkfield tomography\label{subsec:DF-experiment-tomography}}

Tomographic reconstruction, i.e., the transformation of projections
of an object to a volume representation of that object, generally
presumes that all projections correspond to linear combinations of
a static set of scalar volume elements (voxels), which are to be reconstructed.
For samples rotating about an axis parallel to the interferometer's
gratings, as required for tomographic imaging, the distance dependence
of the darkfield contrast inevitably also leads to an apparent orientation
dependence of the signal originating from isotropic scatterers away
from the rotational axis. As has been shown previously, this distance
dependence is not generally negligible and further not independent
of the actual sample properties, which are inherently unknown prior
to reconstruction. Explicit modeling of this dependence in the context
of general iterative tomographic reconstruction methods would therefore
be a non-trivial option. The effects of distance dependence can however
be eliminated to a large degree by symmetric acquisition of projections
from opposing directions, by means of performing full 360° scans:

Due to the linearity -- and thus additivity -- of the darkfield
contrast, individual volume elements (voxels) may be analyzed isolated
without loss of generality. When considering an individual voxel at
a distance $\Delta d$ from the rotational axis (at distance $d_{0}$)
with respect to the optical path, the respective signals from opposing
object orientations are expected to exhibit the following relations
(to first order in $\sigma_{\phi}(d))$:
\begin{align*}
\mu_{\mathrm{DF}}(d(\omega\hphantom{{}+\pi})) & \approx\frac{1}{2}\left[\sigma_{\phi}(d_{0})\,(1+\alpha\,\Delta d_{\omega})\right]^{2}=\mu_{\mathrm{DF}}(d_{0})\,(1+2\alpha\Delta d_{\omega}+\alpha^{2}\Delta d_{\omega}^{2})\\
\mu_{\mathrm{DF}}(d(\omega+\pi)) & \approx\frac{1}{2}\left[\sigma_{\phi}(d_{0})\,(1-\alpha\,\Delta d_{\omega})\right]^{2}=\mu_{\mathrm{DF}}(d_{0})\,(1-2\alpha\Delta d_{\omega}+\alpha^{2}\Delta d_{\omega}^{2})\:,
\end{align*}
with $\alpha$ characterizing the distance dependence as discussed
in the previous Section. The mean signal from opposing orientations
is thus
\[
\frac{1}{2}\left(\mu_{\mathrm{DF}}(d(\omega))+\mu_{\mathrm{DF}}(d(\omega+\pi))\right)=\mu_{\mathrm{DF}}(d_{0})\,(1+\underbrace{\alpha^{2}\Delta d_{\omega}^{2}}_{\mathclap{\le\Delta d_{\omega}^{2}/d_{0}^{2}}})\:,
\]
with $\alpha^{2}\Delta d_{\omega}^{2}$ representing the relative
error with respect to $\mu_{\mathrm{DF}}(d_{0})$. Given that $\alpha\le d_{0}^{-1}$
is expected (cf.\ Eq.~\ref{eq:alpha_D-maximum} and Fig.~\ref{fig:DF-distance-dependence},
lower right), that error is not larger than $\Delta d_{\omega}^{2}/d_{0}^{2}$,
which can be realistically kept below 5\% even for sample diameters
(and thus values of $\Delta d$) of almost up to $d_{0}/2$.

\begin{figure}
\centering{}\includegraphics[width=0.67\textwidth]{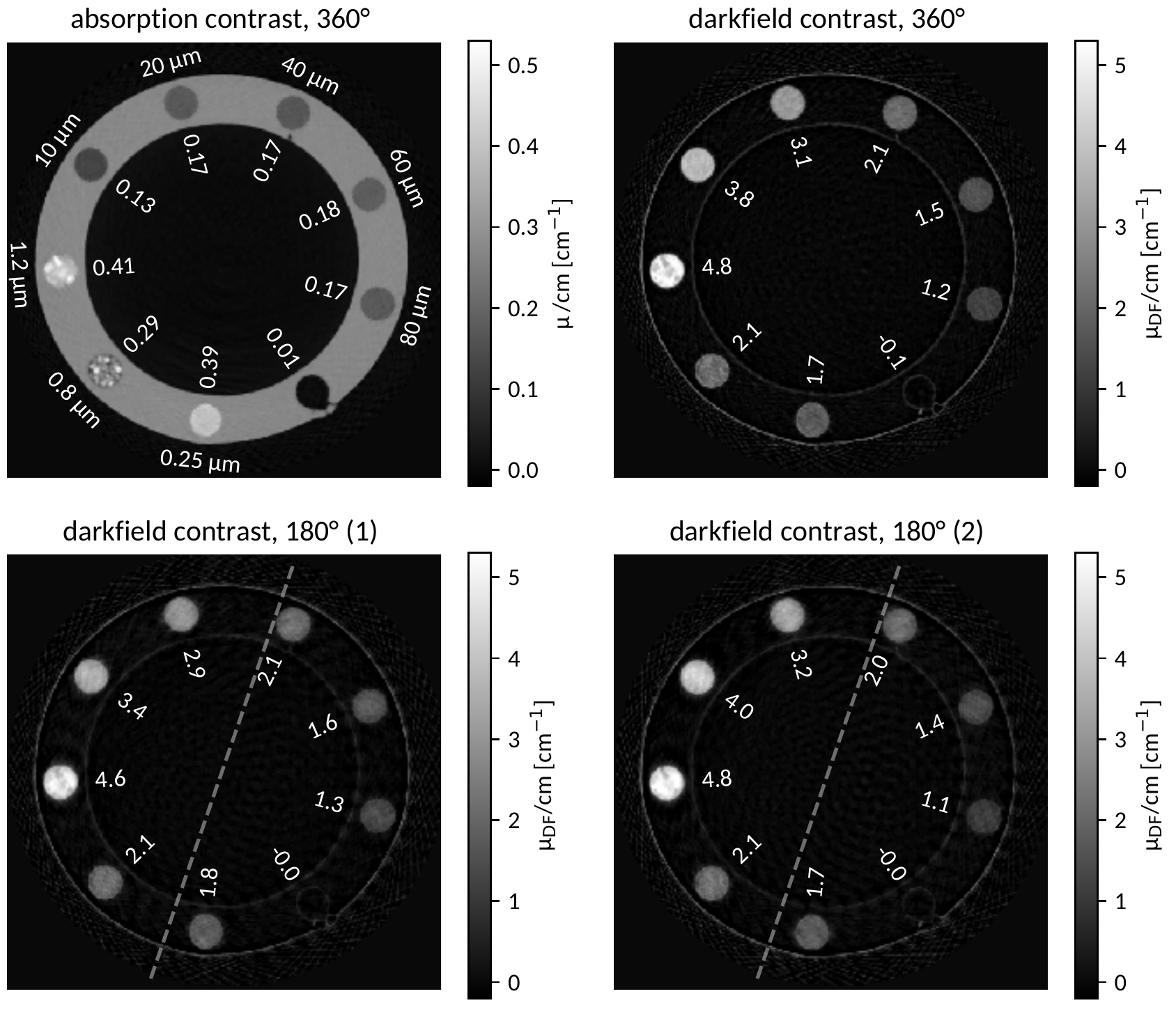}\caption{\label{fig:DF-tomography}Tomographic reconstructions of a cylindrical
sample container comprising nine capillaries filled with spherules
of varying diameter between 0.25\textmu m and 80\textmu m (cf.\ Fig.~\ref{fig:DF-distance-dependence})
from absorption and darkfield projections acquired in parallel beam
geometry. The reconstruction voxel size is 61.6\textmu m, i.e., spherules
are not resolved individually. The respective spherule diameters are
indicated along the outer perimeter, the mean gray values are indicated
along the inner perimeter. Spherules smaller than $5\text{\textmu m}$
are made of glass, while the larger ones consist of PMMA. Variations
in absorption (upper left) within each material class indicate variations
in packing density. The darkfield signal (upper right and lower row)
is further dependent on the size of the spherules. The sample container
itself exhibits darkfield contrast only at edges. The distance dependence
of the darkfield signal is mostly canceled when reconstructing from
the full set of projections (upper right). It has a significant effect
though when reconstructing from 180° subsets (bottom row). Dashed
lines indicate the boundary between over- and underestimations of
the darkfield signal in these cases.}
\end{figure}
\begin{figure}
\centering{}\includegraphics[width=0.5\textwidth]{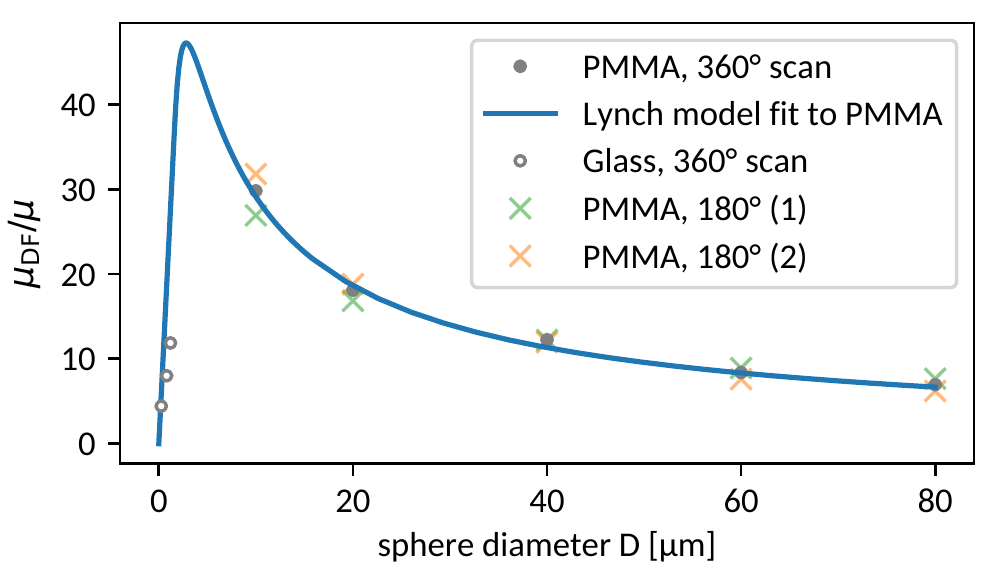}\caption{\label{fig:DF-tomography-quantitative}Evaluation of the darkfield
contrast after tomographic reconstruction (Fig.~\ref{fig:DF-tomography}).
In order to eliminate the volume fractions $f$, the ratio $\mu_{\mathrm{DF}}/\mu$
(darkfield over absorption) is considered. For a given material, variations
in $\mu_{\mathrm{DF}}/\mu$ are now expected to arise solely due to
variations in the sphere diameter $D$. The respective model according
to Lynch (cf.\ Eqs.\ \ref{eq:lynch-darkfield-function}, \ref{eq:lynch-darkfield-function-over-mu})
is compared to the data, assuming $d_{0}=10.7\mathrm{cm}$ as found
previously (Eq.~\ref{eq:DF-distance-dependence-d0}). The absolute
scale is fitted to the data. The good agreement of theory and experiment
support the validity of the Yashiro-Lynch model of darkfield contrast
origination with respect to the predicted structure size dependence.
Further, the quantitative effects of darkfield distance dependence
on short-scan tomographies are shown as well (see also Fig.~\ref{fig:DF-tomography}).}
\end{figure}
Figure~\ref{fig:DF-tomography} shows tomographic reconstructions
of the data previously presented in Figure~\ref{fig:DF-distance-dependence}.
The effects of darkfield distance dependence on tomographic reconstructions
are illustrated by means of reconstructions from two complementary
sets of projections covering an angular range of only 180° each.

A quantitative analysis of the respective gray values is given in
Figure~\ref{fig:DF-tomography-quantitative}. By considering the
absorption normalized darkfield signal $\mu_{\mathrm{DF}}/\mu$, effects
of varying packing density are eliminated, so that variations in darkfield
contrast can be expected to solely arise from differences in material
(Glass vs.\ PMMA) and structure size (spherule diameter). The data
for PMMA is compared to Eq.~\ref{eq:lynch-darkfield-function} (normalized
by the absorption coefficient for PMMA at 35keV) at the correlation
distance $\xi_{0}\approx1.5\text{\textmu m}$ corresponding to the
mean sample--G$_{\text{2}}$ distance $d_{0}\approx10.7\,\mathrm{cm}$
as found previously (Eq.~\ref{eq:DF-distance-dependence-d0}), i.e.,
the distance between the axis of rotation and the analyzer grating
G$_{\text{2}}$:
\begin{equation}
\frac{\mu_{\mathrm{DF}}^{(\mathrm{PMMA})}(D)}{\mu_{\mathrm{PMMA}}^{(35\mathrm{keV)}}}=c_{\mathrm{fit}}\,D\left(1-\sqrt{1-\frac{\xi_{0}^{2}}{D^{2}}}(1+\frac{1}{2}\frac{\xi_{0}^{2}}{D^{2}})+(\frac{\xi_{0}^{2}}{D^{2}}-\frac{1}{4}\frac{\xi_{0}^{4}}{D^{4}})\right)\,,\label{eq:lynch-darkfield-function-over-mu}
\end{equation}
where the proportionality constant fitting the experimental data is
found to be 
\[
c_{\mathrm{fit}}\approx23\,\text{\textmu m}^{-1}\,.
\]
The theoretical expectation (cf.\ Eqs.\ \ref{eq:lynch-darkfield-function}
and \ref{eq:lynch-darkfield-function-over-mu}) evaluates to
\[
\hphantom{c_{\mathrm{fit}}}\mathllap{\frac{\Delta z}{\mu_{\mathrm{PMMA}}^{(35\mathrm{keV)}}}\frac{3\pi^{2}}{\lambda^{2}}\bigl|\Delta\chi\bigr|^{2}=c_{\mathrm{theo}}}\approx\mathrlap{30\,\text{\textmu m}^{-1}}\hphantom{23\,\text{\textmu m}^{-1}\,.}
\]
based on the constants provided in the previous section and $\smash{\smash[t]{\frac{\mu_{\mathrm{PMMA}}^{(35\mathrm{keV)}}}{\Delta z}=0.31\,\mathrm{cm}^{-1}}}$,
as found in the NIST Xcom database assuming a mass density of $1.2\frac{\mathrm{g}}{\mathrm{cm}^{3}}$
for PMMA.

I.e., in addition to the distance dependence discussed previously,
also the size dependence is perfectly consistent with the models given
by Yashiro and Lynch, and can further be verified within tomographic
reconstructions. The absolute quantitative scale found in the experiment
agrees with the theoretic model within a margin of 25\%. 

\subsection{Anisotropic darkfield contrast\label{subsec:aniso-DF-experiments}}

\begin{figure}[p]
\begin{centering}
\includegraphics[width=1\textwidth]{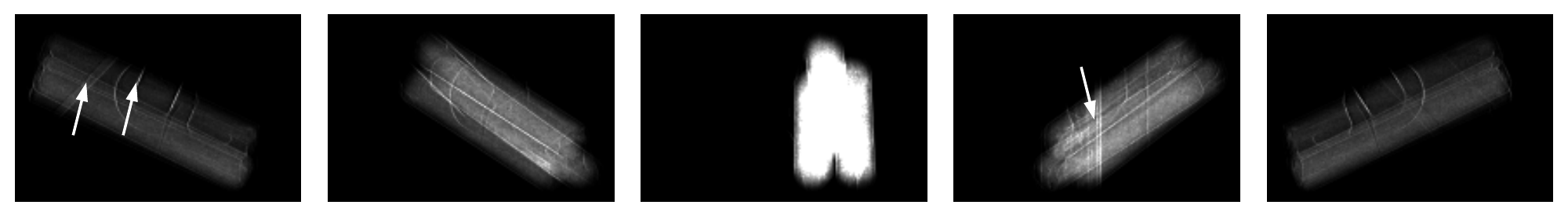}
\par\end{centering}
\centering{}\includegraphics[width=1\textwidth]{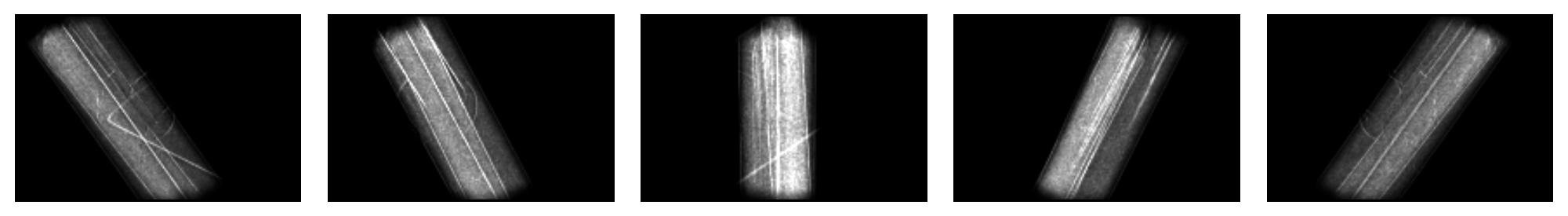}\caption{\label{fig:DF-cfk-image-series}Two darkfield image series of a pack
of three carbon fiber reinforced rods (ca.\ 1cm long) rotating about
the vertical image axis over a range of 180° (0°, 45°, 90°, 135°,
180° from left to right). A sketch is depicted in Fig.~\ref{fig:DF-anisotropy-data-vs-theory}.
The rods are inclined about 65° (top row) and 36° (bottom row) with
respect to the rotational axis. The grating sensitivity is parallel
to the horizontal image axis. White arrows indicate examples of darkfield
signals originating from the sample support structure, which has been
masked outside of the sample silhouette. In the center column, the
projected carbon fiber orientation is perpendicular to the sensitivity
axis. The darkfield contrast is maximal in this case, and for the
top row exceeds the chosen color scale ranging from 0 (black) to 0.7
(white). The center and outermost columns show the isolated effects
of varying scattering cross section (per pixel) and varying autocorrelation
width respectively. Quantitative results are shown in Figure~\ref{fig:DF-anisotropy-data-vs-theory}.}
\end{figure}
\begin{figure}[p]
\begin{centering}
\includegraphics[width=0.63\textwidth]{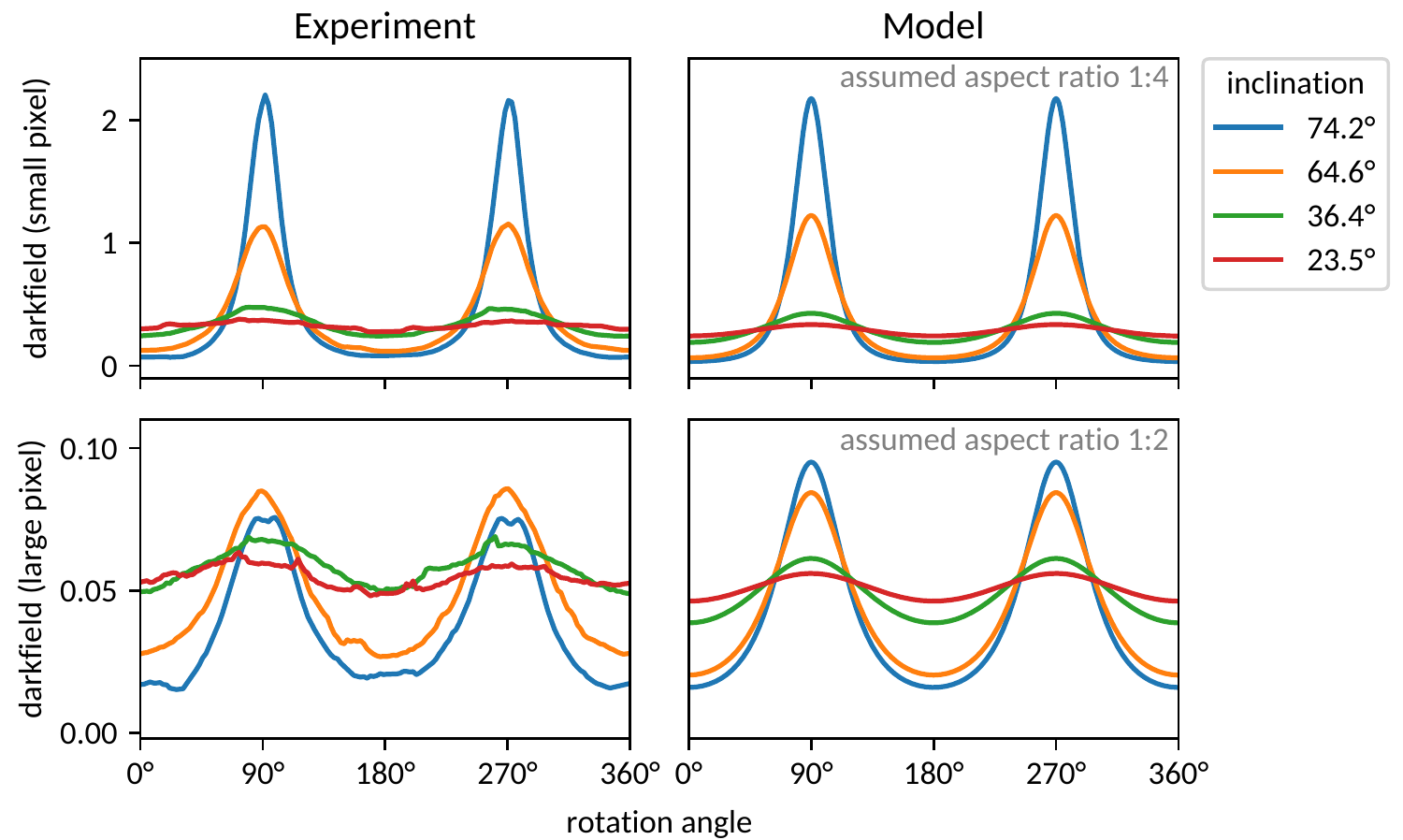}%
\begin{minipage}[b][1\totalheight][t]{0.38\textwidth}%
\begin{flushleft}
\includegraphics[width=1\textwidth]{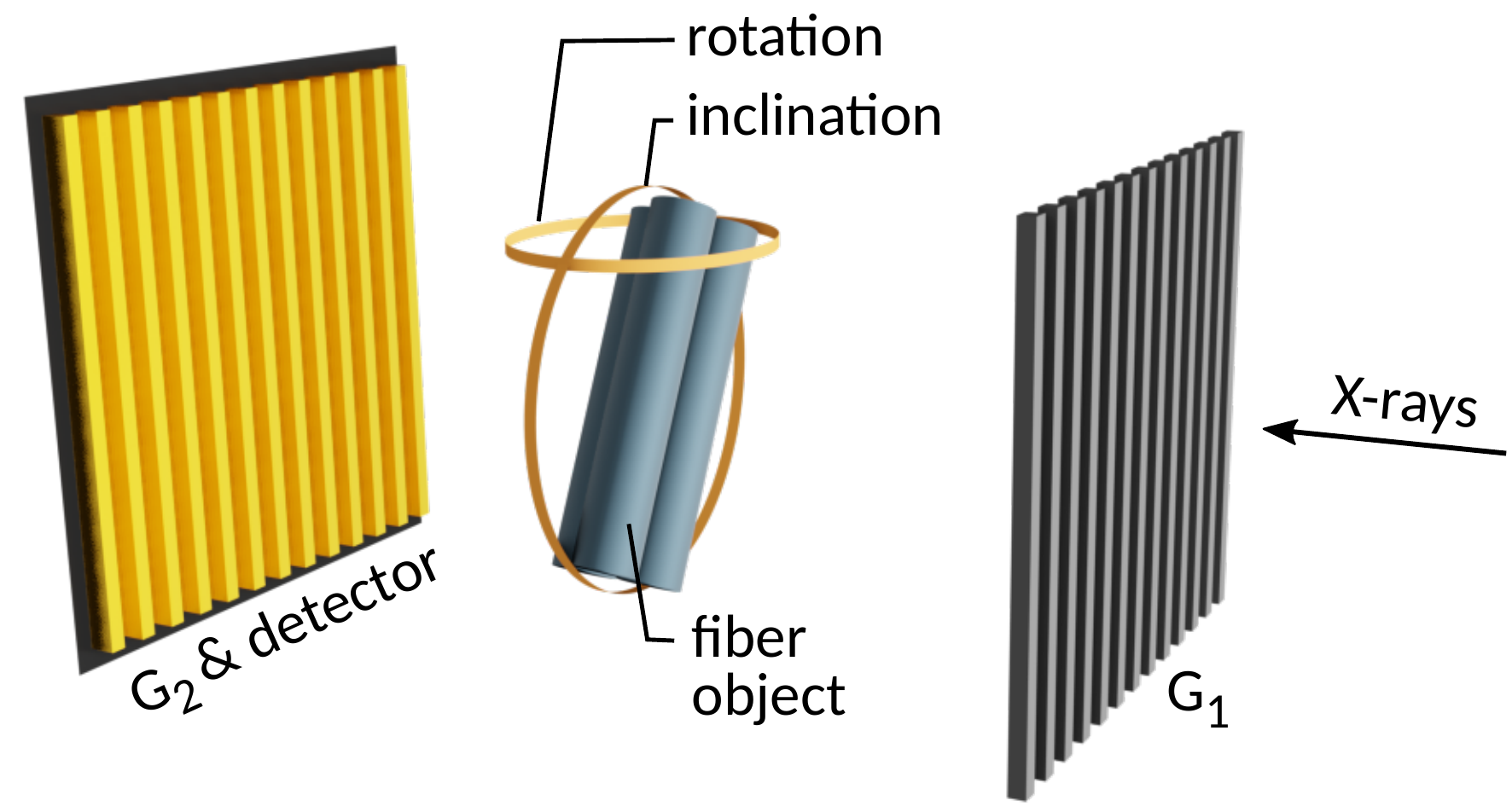}\\
\vspace{.6em}\includegraphics[viewport=19.93464bp 0bp 305bp 110bp,clip,width=0.9\textwidth]{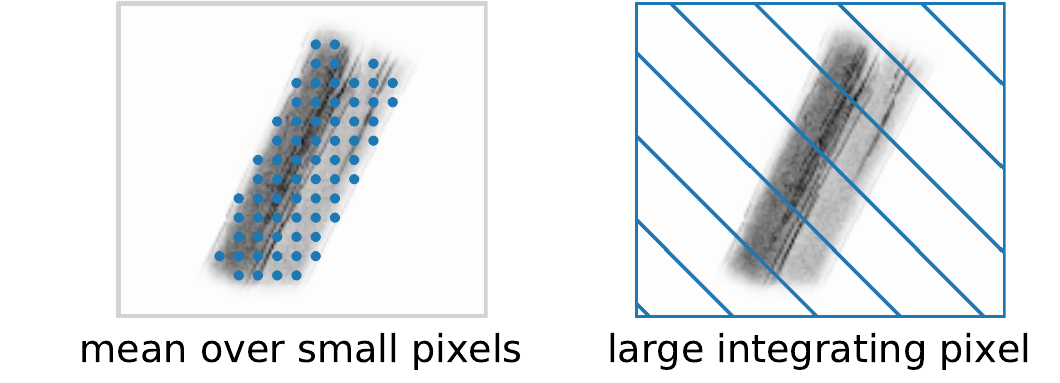}\vspace{.5em}
\par\end{flushleft}%
\end{minipage}
\par\end{centering}
\caption{\label{fig:DF-anisotropy-data-vs-theory}Darkfield signals (negative
logarithm of visibility) for different fiber orientations in the large
pixel and small pixel cases. In the former case, the fibers are always
fully contained within the integration area. Four inclination angles
of the fibers with respect to the rotational axis are considered (cf.\
legend), and a full rotation over 360°  is performed at each inclination.
Experimental data and theoretical model are shown in the left and
right column respectively.  Diameter/length aspect ratios $\sigma_{D}\!:\!\sigma_{L}$
of 1:4 and 1:2 have been assumed for the model data (cf.\ Eqs.~\ref{eq:fiber-bundle-tensor}--\ref{eq:aniso-model-lp-explicit})
shown on the right hand side for the small and large pixel case respectively
in order to approximate the experimental observations on the left
hand side.}
\end{figure}
In order to verify the expected orientation dependence of the darkfield
contrast both with respect to effects of varying autocorrelation width
and scattering cross section, a sample consisting of long carbon fibers
has been imaged at a multitude of orientations in analogy to the experiment
performed by Bayer et al\@.\ \cite{Bayer2012OE}. In contrast to
the latter experiment, the sample is explicitly chosen smaller than
the field of view. The carbon fibers are embedded within three fiber
reinforced plastic rods of about 1cm length and 2mm diameter, and
extend over the full length of the rods. The fiber sample is attached
to a polygonic sample cage (made of UV resin) by means of an acrylic
stand and hot glue. The cage allows to easily vary the inclination
of the carbon fibers with respect to the rotational axis of a tomography
setup, while the acrylic stand centers the sample in the polygonic
cage. Although the support structures are made of non-scattering materials,
a small contribution to the darkfield contrast is generated by their
edges. Figure~\ref{fig:DF-cfk-image-series} shows selected examples.
In order to keep their impact on the following analyses minimal, the
signal of the sample support structures has been masked where possible
(i.e., outside of the sample's silhouette).

The acquired images can be analyzed in two ways: most obviously, the
average darkfield signal per detector pixel over the area of the sample
silhouette may be considered, yielding a signal corresponding to fibers
much longer than the pixel size (analogous to \cite{Bayer2012OE}).
In order to instead reproduce the case of fibers fully contained within
a single integrating pixel, the phase stepping curves' complex amplitudes
 as well as their mean transmission are averaged over the full detector
area prior to the evaluation of visibility and its negative logarithm
(the darkfield signal). The result is then equivalent to that of a
larger integrating detector.

For the comparison of the rotation series acquired at varying fiber
inclinations to the anisotropy model derived in Section~\ref{subsec:aniso-model},
the following mass distribution model of the rotating fibers is used
(cf.\ also Eq.~\ref{eq:density-tensor}):
\begin{equation}
\begin{aligned}\boldsymbol{T}(\omega,\theta) & =\boldsymbol{R}(\omega,\theta)\left[\begin{array}{ccc}
\sigma_{D}^{-2} & 0 & 0\\
0 & \sigma_{L}^{-2} & 0\\
0 & 0 & \sigma_{D}^{-2}
\end{array}\right]\boldsymbol{R}^{T}(\omega,\theta)\\
\boldsymbol{R}(\omega,\theta) & =\left[\begin{array}{ccc}
\hphantom{-{}}\cos\omega & 0 & \sin\omega\\
0 & 1 & 0\\
-\sin\omega & 0 & \cos\omega
\end{array}\right]\left[\begin{array}{ccc}
\cos\theta & -\sin\theta & 0\\
\sin\theta & \hphantom{-{}}\cos\theta & 0\\
0 & 0 & 1
\end{array}\right]
\end{aligned}
\label{eq:fiber-bundle-tensor}
\end{equation}
with $\sigma_{D}$ and $\sigma_{L}$ denoting standard deviations
characterizing the effective (with respect to darkfield effects) mean
diameter and length of the fiber bundle and $\omega$ and $\theta$
describing its orientation in terms of rotation and inclination as
sketched in Figure~\ref{fig:DF-anisotropy-data-vs-theory}.

For the first case considering small detector pixels (indicated by
the subscript ``$\mathrm{sp}$''), the following is expected from
the previous theoretic derivations (cf.\ Eq.~\ref{eq:aniso-model-sp}):
\begin{align}
-\ln(v_{\mathrm{sp}})\appropto\frac{1}{T_{zz}}(T_{\mathrm{xx}}-\frac{T_{\mathrm{xz}}^{2}}{T_{\mathrm{zz}}}) & \,\propto\hphantom{\frac{1}{\sigma_{D}}}\frac{\quad\!\cos^{2}\theta+\frac{\sigma_{D}^{2}}{\sigma_{L}^{2}}\sin^{2}\theta}{(\cos^{2}\omega+(\cos^{2}\theta+\frac{\sigma_{D}^{2}}{\sigma_{L}^{2}}\sin^{2}\theta)\sin^{2}\omega)^{2}}\:.\label{eq:aniso-model-sp-explicit}\\
\intertext{\text{For the case of a single large detector (“\ensuremath{\mathrm{lp}}”) integrating over the full extent of all fibers,}}-\ln(v_{\mathrm{lp}})\appropto\frac{1}{\sqrt{T_{zz}}}(T_{\mathrm{xx}}-\frac{T_{\mathrm{xz}}^{2}}{T_{\mathrm{zz}}}) & \,\propto\frac{1}{\sigma_{D}}\frac{\cos^{2}\theta+\frac{\sigma_{D}^{2}}{\sigma_{L}^{2}}\sin^{2}\theta}{(\cos^{2}\omega+(\cos^{2}\theta+\frac{\sigma_{D}^{2}}{\sigma_{L}^{2}}\sin^{2}\theta)\sin^{2}\omega)^{3/2}}\label{eq:aniso-model-lp-explicit}
\end{align}
 is expected in contrast (cf.\ Eq.~\ref{eq:aniso-model-lp}).

Figure~\ref{fig:DF-anisotropy-data-vs-theory} shows a respective
comparison of experimental data and theoretic model. While the individual
carbon fibers contained in the considered sample are expected to have
an aspect ratio of about $10^{3}$ (ca.\ $10^{-2}\mathrm{m}$ length
at ca.\ $10^{-5}\mathrm{m}$ diameter), the observed signal is best
reproduced with aspect ratios $\sigma_{L}/\sigma_{D}$ of $4$ and
$2$ in the small and large pixel case respectively. The general discrepancy
between the extreme aspect ratio of individual fibers and the deduced
aspect ratios of the fiber ensemble is expected to arise from a finite
distribution width of fiber orientations. Such variations in orientation
are generated whenever fibers are bent or not perfectly aligned parallel,
which is especially expected among the three separate rods constituting
the sample. Similarly, the reduced aspect ratio found for the case
of the integrating detector (as compared to the small pixel case)
might be attributed to the larger ensemble of fibers considered simultaneously
in that case. 

While the present data doesn't allow further microscopic analyses
of the observed aspect ratios, the observed orientation dependence
with respect to rotations and inclinations is in good agreement with
the theoretic expectation despite the considerable number of first
order approximations that have been made towards the derivation of
Eqs.~\ref{eq:aniso-model-sp-explicit}--\ref{eq:aniso-model-lp-explicit}
(cf.\ Section~\ref{sec:aniso-df}). First of all, both the dependence
on changes in scattering cross section and in autocorrelation width
are reproduced, although the integrating pixel case appears to be
more susceptible to imperfections of the sample. The scattering cross
section dependence is found to be -- as expected by Eq.~\ref{eq:aniso-model-sp-explicit}
-- considerably more pronounced in the case of objects exceeding
the pixels' integration area. Moreover, the narrowly peaked rotation
angle dependence for strongly inclined fibers is reproduced, which
can be attributed to the influence of the off-diagonal term $T_{\mathrm{xz}}^{2}/T_{zz}$
originating from inclinations of the anisotropic mass distribution
with respect to the optical axis (cf.\ Figure~\ref{fig:DF-scattering-cross-section}
left).

\section{Conclusions}

The origination of darkfield contrast has been discussed extensively
in the past from various points of view. The existing variety of approaches
can be shown to be largely consistent with the wave optical derivations
by Yashiro and Lynch, which on the other hand provide the crucial
link to Fresnel optics, which allows for a well founded extension
of the existing models to cone beam geometries based on the Fresnel
scaling relation. Following the argumentation of Yashiro and Lynch,
the results can furthermore be extended to anisotropic scatterers.
A complete model considering both the anisotropy of the autocorrelation
width as well as the scattering cross section is derived, as required
for the description of arbitrarily oriented fibers (as opposed to
anisotropy considerations solely within a planar sample perpendicular
to the optical axis). All results are supported by experiments, with
particular focus on the demands of quantitative tomography. Additional
effects relevant to quantitative darkfield interpretations that have
not been explicitly considered here, yet shall not be left unmentioned,
include influences of beam polychromaticity \cite{Kaeppler2014,Pelzer2016,Yashiro2018},
as well as discontinuities at material boundaries and higher order
optical effects affecting the periodicity of the Talbot pattern \cite{Wolf2015}.

\section*{Acknowledgments }

The authors gratefully acknowledge the beamtime granted at the European
Synchrotron Radiation Facility ESRF (experiment MA4346), as well as
the contributors to the employed grating interferometer setup. M.
Olbinado and A. Rack are acknowledged for the excellent user support
at ID19, and D. Müller, M. Ullherr and M. Seitz for their kind support
during the preparation and performance of the experiments. C. Fella
is acknowledged for providing the required freedom to finish this
work. Funding is acknowledged from the Bavarian State Ministry of
Economic Affairs, Infrastructure, Transport and Technology which supported
the project group ”Nano-CT Systems for Material Characterization'',
and the LEE-BED EU project no.\ 814485. Furthermore, the open source
projects Debian, Python, SciPy, PyOpenCL, matplotlib, Jupyter, Blender,
LyX and TeX Live (and many more behind the scenes) are acknowledged,
without which the present manuscript would not have been possible. 

\paragraph*{Author's Contributions: }

J.G. conceived and performed the study, did the literature research,
worked out the theory, wrote the image processing software, performed
the data analysis and visualization and wrote the manuscript. J.G.,
A.B. and S.Z conceived, prepared and performed the experiments (with
support by M.O., A.R., D.M, M.U., M.S.) and critically revised the
manuscript. R.H. supervises the PhD project of J.G., secured funding
and critically revised the manuscript. All authors approved the final
version of this manuscript.

\end{document}